\documentclass[Afour,sageh,times]{sagej}

\usepackage{moreverb,url}
\usepackage{algorithm}
\usepackage{algorithmic}
\usepackage[shortlabels]{enumitem}
\usepackage[subrefformat=parens]{subcaption}

\usepackage[colorlinks,bookmarksopen,bookmarksnumbered,citecolor=red,urlcolor=red]{hyperref}

\newcommand\BibTeX{{\rmfamily B\kern-.05em \textsc{i\kern-.025em b}\kern-.08em
T\kern-.1667em\lower.7ex\hbox{E}\kern-.125emX}}

\def\volumeyear{2016}

\begin{document}

\runninghead{Uchino, Ozaki, and Imamura}

\title{Performance Enhancement\\of the Ozaki Scheme\\on Integer Matrix Multiplication Unit}

\author{Yuki Uchino\affilnum{1}, Katsuhisa Ozaki\affilnum{2}, and Toshiyuki Imamura\affilnum{1}}

\affiliation{\affilnum{1}RIKEN Center for Computational Science, Japan\\
\affilnum{2}Department of Mathematical Sciences, Shibaura Institute of Technology, Japan}

\corrauth{Yuki Uchino, 
	7-1-26 Minatojima-minami-machi, Chuo-ku, Kobe, Hyogo 650-0047, Japan}

\email{yuki.uchino.fe@riken.jp}

\begin{abstract}
This study was aimed at simultaneously achieving sufficient accuracy and high performance for general matrix multiplications. Recent architectures, such as NVIDIA GPUs, feature high-performance units designed for low-precision matrix multiplications in machine learning models, and next-generation architectures are expected to follow the same design principle. The key to achieving superior performance is to fully leverage such architectures.
The Ozaki scheme, a highly accurate matrix multiplication algorithm using error-free transformations, enables higher-precision matrix multiplication to be performed through multiple lower-precision matrix multiplications and higher-precision matrix additions. Ootomo et al. implemented the Ozaki scheme on high-performance matrix multiplication units with the aim of achieving both sufficient accuracy and high performance.
This paper proposes alternative approaches to improving performance by reducing the numbers of lower-precision matrix multiplications and higher-precision matrix additions. Numerical experiments demonstrate the accuracy of the results and conduct performance benchmarks of the proposed approaches. These approaches are expected to yield more efficient results in next-generation architectures.

\end{abstract}


\maketitle

\section{Introduction}
\label{sec:Introduction}
The field of machine learning, including AI, is evolving daily, and the scale and complexity of machine learning models are continually increasing. 
Recent architectures are designed to process these models rapidly and with high energy efficiency. 
For machine learning models, matrix multiplications using low-precision floating-point systems and integers are essential. 
Therefore, recent architectures are equipped with high-performance low-precision floating-point and integer matrix multiplication units, and high-performance mixed-precision matrix multiplication units that leverage these capabilities.
A prime example of this is the NVIDIA Tensor Cores (see~\cite{tensorcore}).
Table~\ref{tab:Specifications} shows the specifications of the NVIDIA GPUs equipped with Tensor Core technology.
Note that the specification of FP16 TC on RTX 4090 (165 TFLOPS) is for FP16 input and FP32 output, and the specifications for the H100 and H200 are the same as those for the GH200.
In the future, numerical computation algorithms that leverage the performance of cutting-edge architectures will be essential. 
This study focused on researching high-performance matrix multiplication algorithms that maximize the potential of the latest architectures.

\begin{table}[htb]
	\centering
	\caption{Specifications in TFLOPS/TOPS of NVIDIA GPUs for dense data~\cite{tensorcore}}
	\label{tab:Specifications}
	\begin{tabular}{l@{ }l@{\ \ }c@{\ \ }c@{\ \ }c@{\ \ }c@{\ \ }c}
		\hline
		&    & RTX 4090 & A100 & GH200 & B200 & GB200 \\ \hline
		FP64 &    &  1.29   & 9.7  &    34     &  40  & 90 \\
		FP64 & TC &    --    & 19.5 &    67     &  40  & 90 \\
		FP32 &    &  82.6   & 19.5 &    67     &  80  & 180 \\
		TF32 & TC &  82.6   & 156  &    494    & 1100 & 5000 \\
		BF16 & TC &   165   & 312  &    989    & 2250 & 5000 \\
		FP16 & TC & 165 & 312  &    989    & 2250 & 5000 \\
		INT8 & TC &   661   & 624  &   1979    & 4500 & 10000 \\
		FP8  & TC &   661   &  --   &   1979    & 4500 & 10000 \\
		FP6  & TC &    --    &  --   &     --     & 4500 & 10000 \\
		INT4 & TC &  1321   &  --   &     --     &  --   & -- \\
		FP4  & TC &         &  --   &     --     & 9000 & 20000 \\ \hline
	\end{tabular}
\end{table}

A highly accurate matrix multiplication scheme via the error-free transformation of matrix products was proposed in~\cite{ozaki2012error}.
The scheme is called the Ozaki scheme and it converts a matrix product into a sum of multiple matrix products.
It is also possible to convert a matrix product into a sum of lower-precision matrix products to take full advantage of the immense computational power of recent architectures.
In order to compute $D\gets AB$ for higher-precision matrices $A \in \mathbb{R}^{m \times n}$ and $B \in \mathbb{R}^{n \times p}$, the Ozaki scheme using lower/mixed-precision, provided in \cite{mukunoki2020}, is constructed as the following four steps:
\begin{enumerate}[(i),ref={(\roman*)}]
	\item \label{item1} Extract the lower-precision matrices $A_1,A_2,\dots,A_k$ from the higher-precision matrix $A$, where $k$ is specified by the user, shifting $A_i$ to prevent overflow and underflow at lower-precision arithmetic.
	\item \label{item2}Extract $k$ lower-precision matrices $B_i$ from $B$ in the similar way as $A_i$.
	\item \label{item3}Compute $A_iB_j$ for $i+j <= k+1$ using lower/mixed-precision arithmetic.
	\item \label{item4}Reverse (shift) $A_iB_j$ and accumulate the results into $D$ using higher-precision arithmetic.
\end{enumerate}
When emulating the GEMM routine, the following fifth step is also performed:
\begin{enumerate}[(i),ref={(\roman*)}]\setcounter{enumi}{4}
	\item \label{item5}Compute $C \gets \alpha D + \beta C$ for scalars $\alpha,\beta$ and a matrix $C$.
\end{enumerate}

\begin{figure}[htb]
	\centering
	\includegraphics[width=\hsize]{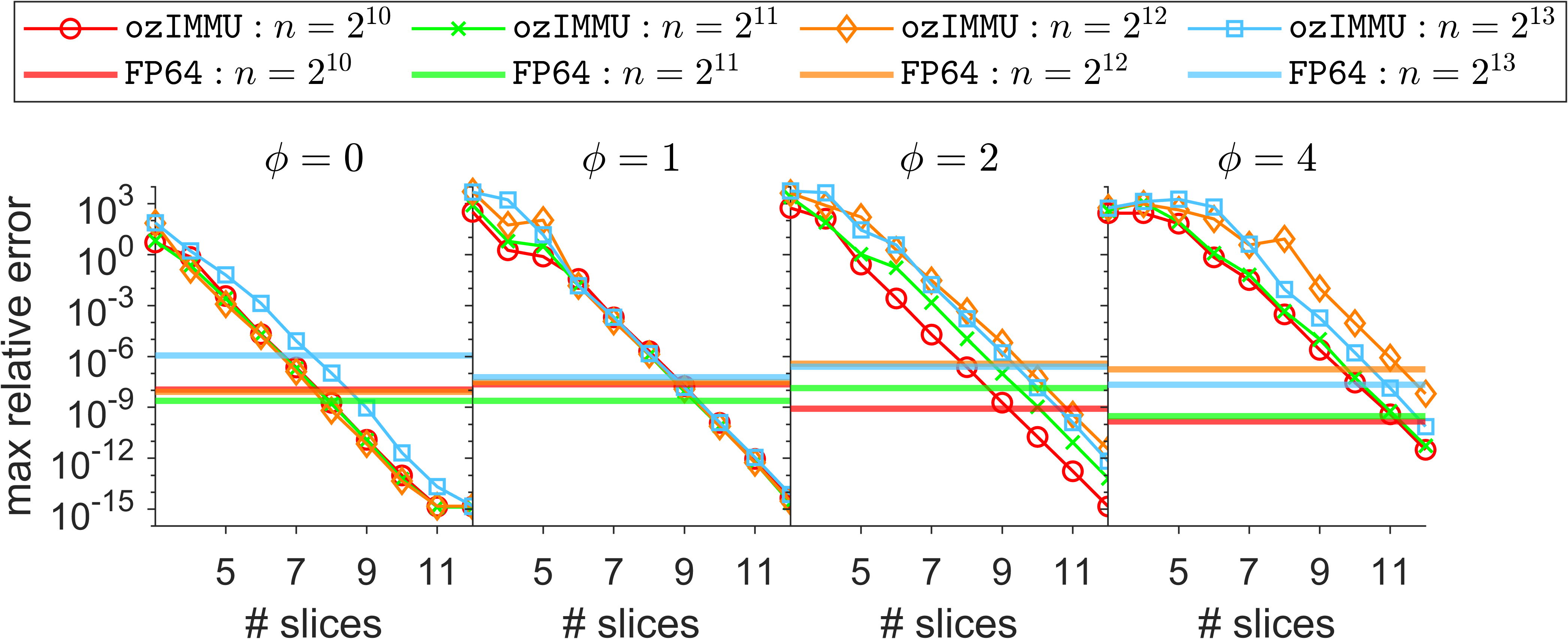}
	
	\caption{Accuracy of ozIMMU. Matrix $A$ has entries $a_{ij} := (U_{ij}-0.5)\cdot \exp(\phi\cdot N_{ij})$, where $U_{ij} \in (0,1)$ are uniformly distributed and $N_{ij}$ are drawn from standard normal distribution for $1 \le i,j \le m$ and $m=n=p$. Matrix $B$ is composed similarly.}
	\label{fig:accuracy_ozIMMU.png}
\end{figure}

Ootomo et al. implemented the Ozaki scheme using the INT8 Tensor Cores and evaluated the performance for emulating DGEMM, an FP64 matrix multiplication routine, as reported in~\cite{ootomo2024dgemm}.
The Ozaki scheme implemented by Ootomo is named the ``Ozaki Scheme on Integer Matrix Multiplication Unit'' (ozIMMU), and the code is available in~\cite{ozIMMU}.
Figure~\ref{fig:accuracy_ozIMMU.png} shows the accuracy of ozIMMU.
Herein, $\phi$ specifies the tendency of the difficulty in terms of the accuracy of matrix multiplications.
At least 7 or 8 slices are required to obtain sufficiently accurate results, even for well-conditioned matrix multiplications.
As $\phi$ increases, more slices are required to obtain sufficient accuracy.
Ootomo's implementation for splitting matrices offers bit masking.
Thus, the extracted matrices $A_i, B_i$ may not be optimal and the splitting method can be improved (see Section~\ref{subsec:Proposed1} for details).
Improvement of the splitting method should contribute to improving the accuracy of results.

\begin{figure}[htb]
	\centering
	\includegraphics[width=\hsize]{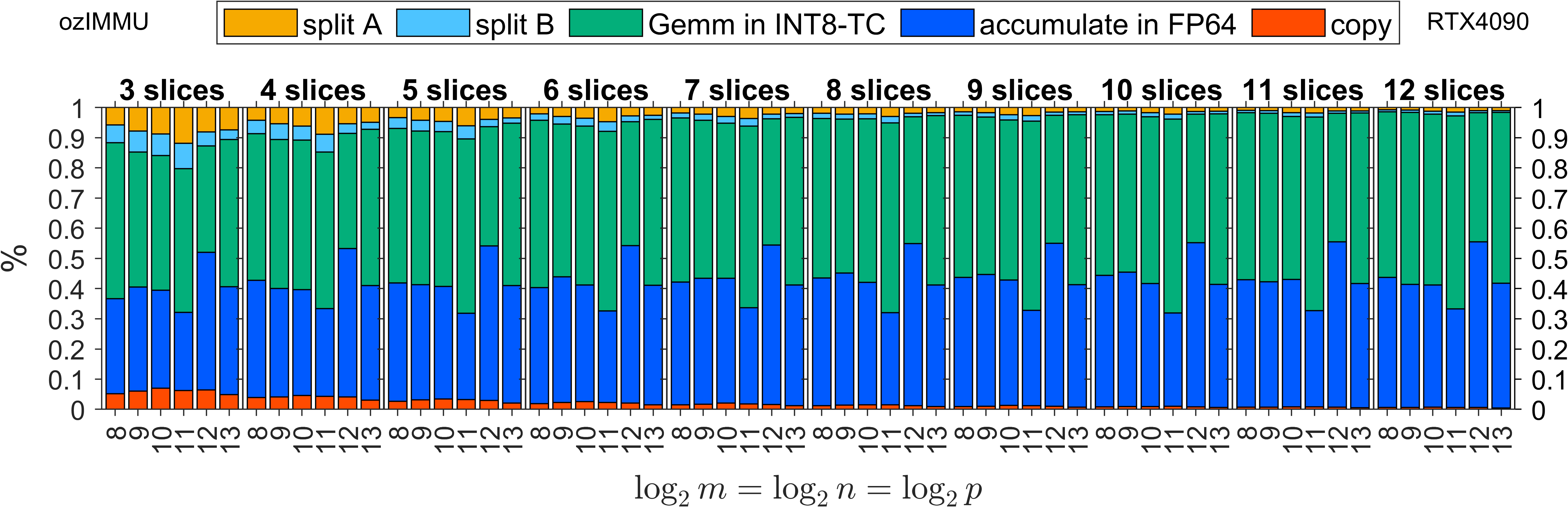}
	\caption{Time breakdown of ozIMMU on NVIDIA GeForce RTX 4090}
	\label{fig:time_breakdown_ozIMMU_RTX4090.png}
\end{figure}

\begin{figure}[htb]
	\centering
	\includegraphics[width=\hsize]{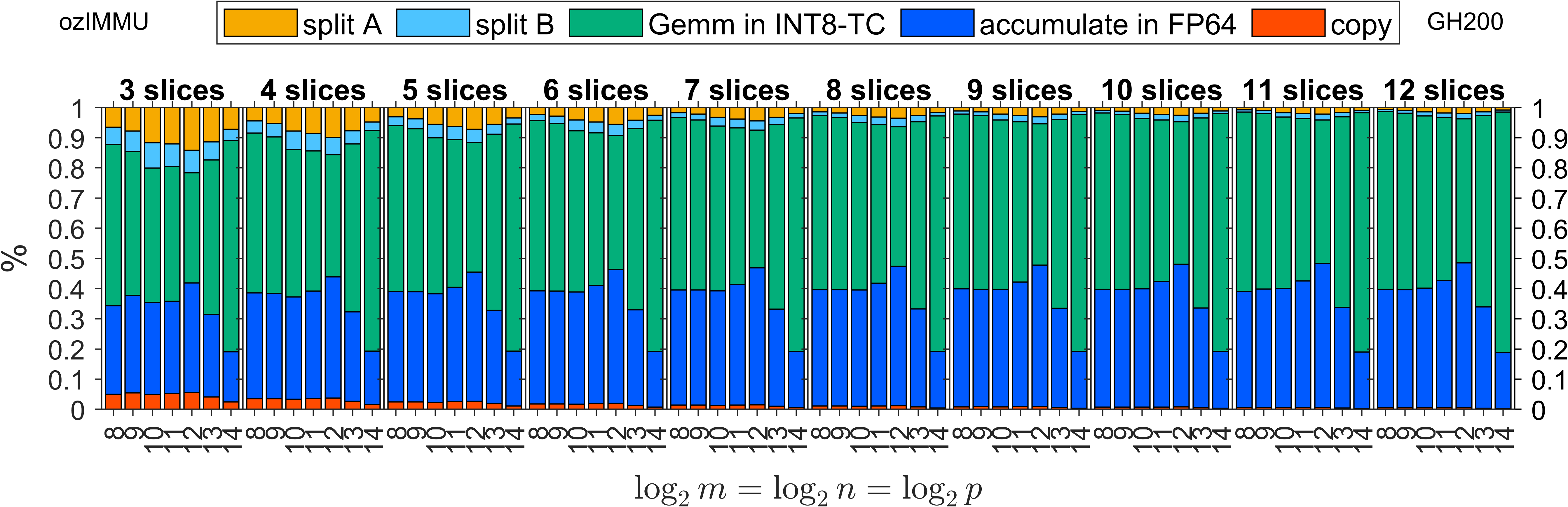}
	\caption{Time breakdown of ozIMMU on NVIDIA GH200 Grace Hopper Superchip}
	\label{fig:time_breakdown_ozIMMU_gh200.png}
\end{figure}

Figures~\ref{fig:time_breakdown_ozIMMU_RTX4090.png} and \ref{fig:time_breakdown_ozIMMU_gh200.png} show the time breakdown of ozIMMU for double-precision matrices $A, B$, defined in~\cite{ieee754}, on GeForce RTX 4090 and GH200 Grace Hopper Superchip, respectively.
Note that ``split A'', ``split B'', ``Gemm in INT8-TC'', ``accumulation in FP64'', and ``copy'' correspond to the steps \ref{item1}, \ref{item2}, \ref{item3}, \ref{item4}, and \ref{item5} in the Ozaki scheme for emulating the GEMM routine, respectively.
It can be seen that the computing time for the accumulation of matrix products in FP64 is not negligible even though the computation cost is $\mathcal{O}(mp)$ operations.
This is because the performance of the INT8 Tensor Core is nearly 512 times and 60 times higher than that of FP64 on the RTX 4090 and GH200, respectively.
Because this ratio increases significantly on the B200 and future architectures, accelerating the accumulation process is critical for the Ozaki scheme.

In this paper, we propose new implementation methods for accelerating the accumulation of the Ozaki scheme using the INT8 Tensor Core.
In addition, an alternative splitting method is also applied to improve the accuracy of results.
For the number of slices, the splitting method contributes to obtaining more accurate results than those of Ootomo's ozIMMU.
The remainder of this paper is organized as follows: 
Section~\ref{sec:Previous study} overviews previous studies about the Ozaki scheme using an integer matrix multiplication unit;
Section~\ref{sec:Proposed methods} presents the proposed method for accelerating the accumulation and optimizing the splitting method;
Section~\ref{sec:Numerical examples} shows numerical results to illustrate the efficiency of the proposed methods;
Section~\ref{sec:Rounding error analysis} provides a rounding error analysis of the Ozaki scheme with the proposed methods;
and Section~\ref{sec:Conclusion} presents final remarks.

\section{Previous study}
\label{sec:Previous study}

In this section, we briefly summarize previous studies.
Let $u$ be the relative error unit, e.g., $u = 2^{-53}$ for double-precision floating-point numbers defined in IEEE~754.
Define $\mathbb{F}$ as a set of binary floating-point numbers with $u$. 
Let $O_{m,n}$ be an $m \times n$ zero matrix.
Suppose that $A \in \mathbb{F}^{m \times n}$ and $B \in \mathbb{F}^{n \times p}$.
Ozaki et al. proposed an error-free transformation of a matrix product $AB$ in~\cite{ozaki2012error}.
For a user-specified constant $k \in \mathbb{N}$, the Ozaki scheme transforms each of $A$ and $B$ into $k+1$ matrices
\begin{align*}
    A &=: A_1 + A_2 + \dots + A_{k} + V_{k},\\
    B &=: B_1 + B_2 + \dots + B_{k} + W_{k},
\end{align*}
where $A_i \in \mathbb{F}^{m \times n}$, $B_i \in \mathbb{F}^{n \times p}$, 
\begin{equation}\label{def:VW}
V_i := A - \sum_{j=1}^{i} A_j,\quad \text{and}\quad
W_i := B - \sum_{j=1}^{i} B_j,
\end{equation}
such that $V_i \in \mathbb{F}^{m \times n}$ and $W_i \in \mathbb{F}^{n \times p}$ for $i \le k$, and $|(A_s)_{ij}| \ge |(A_{s+1})_{ij}|$ holds if $(A_s)_{ij} \neq 0$ for $s=1,\dots,k-2$, and $B_s$ satisfies the corresponding relation.
If the technique provided in~\cite{minamihata2016improved} is used, then $|(A_s)_{ij}| > |(A_{s+1})_{ij}|$ holds when $(A_s)_{ij} \neq 0$.

Let $\mathbb{F}_{N}$ be a set of $N$-bit
binary floating-point numbers.
For the Ozaki scheme using the FP16 Tensor Core with accumulation in FP32 and outputs in FP32, the mixed-precision splitting method with shifting to prevent overflow and underflow was utilized in~\cite{mukunoki2020}.
For $A \in \mathbb{F}_{64}^{m \times n}$ and $B \in \mathbb{F}_{64}^{n \times p}$, the splitting method produces
\begin{align}
	A &=: \mathrm{diag}(\mu'^{(1)}) A'_1  + \dots + \mathrm{diag}(\mu'^{(k)}) A'_{k} + V_{k},\label{eq:splitA'}\\
	B &=: B'_1 \mathrm{diag}(\nu'^{(1)}) + \dots + B'_k \mathrm{diag}(\nu'^{(k)}) + W_{k}\label{eq:splitB'}
\end{align}
with double-precision shift values $\mu'^{(i)} \in \mathbb{F}_{64}^m$, $\nu'^{(i)} \in \mathbb{F}_{64}^p$ and half-precision matrices $A'_i \in \mathbb{F}_{16}^{m \times n}$, $B'_i\in \mathbb{F}_{16}^{n \times p}$ for $i=1,\dots,k$.
Algorithm~\ref{alg:OzakiSplit-mukunoki} represents the splitting method for obtaining \eqref{eq:splitA'} using Minamihata's technique.
Equation \eqref{eq:splitB'} can be obtained by transposing the results of Algorithm~\ref{alg:OzakiSplit-mukunoki} executed for $B^T$.
Algorithm~\ref{alg:OzakiSplit-mukunoki} can be described as a loop that is executed until $s=k$ for simplicity; however, the loop terminates when $A=O_{m,n}$ in practice.
Note that the binary logarithm is used at the \ref{alg:OzakiSplit-mukunoki-3}rd line in Algorithm~\ref{alg:OzakiSplit-mukunoki}; however, the calculation using the binary logarithm occasionally returns erroneous results. Therefore, it is better to use a calculation method without the binary logarithm, such as a bitwise operation or a technique leveraging rounding error in floating-point arithmetic as follows:
\[
\mu'^{(s)}_i \gets u^{-1}\alpha_i + (1-u^{-1})\alpha_i,
\]
where $\alpha_i := \max_j |a_{ij}|$.
This technique was developed by Rump and provided in ~\cite{ozaki2013generalization}.

\begin{algorithm}[htb]
	\caption{Mixed-precision splitting method from~\cite{mukunoki2020} for Ozaki scheme between FP64 and FP16 using floating-point arithmetic in round-to-nearest-even mode with Minamihata's technique from~\cite{minamihata2016improved}\label{alg:OzakiSplit-mukunoki}}
	\begin{algorithmic}[1]
		\REQUIRE $A \in \mathbb{F}_{64}^{m \times n}$, $k \in \mathbb{N}$
		\ENSURE $A'_s \in \mathbb{F}_{16}^{m \times n}$, $\mu'^{(s)} \in \mathbb{F}_{64}^m$, $s=1,\dots,k$
		\STATE $\beta \gets \lceil (29 - \log_2 n )/2 \rceil$
		\FOR{$s=1,\dots,k$}
		\STATE\label{alg:OzakiSplit-mukunoki-3} $\mu'^{(s)}_i \gets 2^{\lceil \log_2 \max_j |a_{ij}| \rceil}$ for $i=1,\dots,m$ 
		\STATE $\sigma_i \gets 0.75 \cdot \mu'^{(s)}_i \cdot 2^{\beta}$ for $i=1,\dots,m$
		\STATE $(A_s)_{ij} \gets (a_{ij} + \sigma_i) - \sigma_i$  $\forall i,j$
		\hfill\COMMENT{Extract in round-to-nearest-even mode}
		\STATE $(A'_s)_{ij} \gets \mathrm{FP16}((\mu'^{(s)}_i)^{-1}\cdot (A_s)_{ij})$  $\forall i,j$
		\hfill
		\COMMENT{Convert to FP16}
		\STATE $a_{ij} \gets a_{ij}-(A_s)_{ij}$  $\forall i,j$
		\ENDFOR
	\end{algorithmic}
\end{algorithm}

After splitting matrices as in \eqref{eq:splitA'} and \eqref{eq:splitB'}, 
\[\begin{aligned}
AB 
&= \sum_{s+t \le k+1} \mathrm{diag}(\mu'^{(s)}) A'_sB'_t \mathrm{diag}(\nu'^{(t)})\\
&+ \sum_{s+t > k+1} \mathrm{diag}(\mu'^{(s)}) A'_sB'_t \mathrm{diag}(\nu'^{(t)})\\
&+ \sum_{s=1}^k \mathrm{diag}(\mu'^{(s)}) A'_s W_{k}\\
&+ \sum_{s=1}^k V_{k} B'_s \mathrm{diag}(\nu'^{(s)})\\
&+ V_{k}W_{k}
\end{aligned}
\]
holds and there is no rounding error in $A'_sB'_t$ for $1 \le s,t \le k$ on the FP16 Tensor Core.
In~\cite{mukunoki2020}, the approximation of $AB$ was computed only using matrix multiplications with the FP16 Tensor Core, as 
\[
AB \approx \sum_{s+t \le k+1} \mathrm{diag}(\mu'^{(s)}) A'_sB'_t \mathrm{diag}(\nu'^{(t)})
\]
shown in Algorithm~\ref{alg:Ozakimul-mukunoki}.
This method was referred to as the ``Fast Mode'' in~\cite{mukunoki2020}.
For larger $k$, $|(W_{k})_{ij}|$ and $|(V_k)_{ij}|$ are smaller and the approximation is more accurate.

\begin{algorithm}[htb]
	\caption{Mixed-precision matrix multiplication method for Ozaki scheme using FP16 Tensor Core in fast mode from~\cite{mukunoki2020}\label{alg:Ozakimul-mukunoki}}
	\begin{algorithmic}[1]
		\REQUIRE $A'_s \in \mathbb{F}_{16}^{m \times n}$, $\mu'^{(s)} \in \mathbb{F}_{64}^m$, $B'_s \in \mathbb{F}_{16}^{n \times p}$, $\nu'^{(s)} \in \mathbb{F}_{64}^n$, $s=1,\dots,k$
		\ENSURE $C \in \mathbb{F}_{64}^{m \times p}$
		\STATE $C \gets O_{m,p}$
  \hfill\COMMENT{$C' \in \mathbb{F}^{m \times p}_{32}$}
		\FOR{$g = 2,\dots,k+1$}
		\FOR{$s = 1,\dots,g-1$}
		\STATE $C' \gets A'_sB'_{g-s}$
		\hfill\COMMENT{Compute using \texttt{GEMM} on FP16 Tensor Core}
		\STATE $C \gets C + \mathrm{diag}(\mu'^{(s)})\mathrm{FP64}(C')\mathrm{diag}(\nu'^{(t)})$
		\hfill\COMMENT{Compute in FP64}
		\ENDFOR
		\ENDFOR
	\end{algorithmic}
\end{algorithm}

Let $\mathbb{I}_{N}$ be a set of $N$-bit signed integers.
It holds that $-2^{N-1} \le i \le 2^{N-1}-1$ for all $i \in \mathbb{I}_{N}$.
Remember that no error occurs in integer arithmetic, barring overflow.
For the Ozaki scheme using the INT8 Tensor Core with accumulation in INT32, the mixed-precision splitting method via bit masking with shifting is offered in \cite{ootomo2024dgemm} and provided in~\cite{ozIMMU}.
Algorithm~\ref{alg:OzakiSplit-ootomo} shows the splitting method.
Let 
\begin{equation}
	\beta := \min\left( 7, \left\lfloor \frac{31 - \log_2 n}{2} \right\rfloor \right).
	\label{eq:beta}    
\end{equation}
Assume that $\beta \ge 1$, i.e., $n \le 2^{29}$.
That splitting method produces
\begin{align}
	A &=: \mathrm{diag}(\mu'') (2^{-\beta +1}A''_1 + \dots + 2^{-k \beta +1}A''_{k}) + V_{k},\label{eq:splitA''}\\
	B &=: (2^{-\beta +1}B''_1 + \dots + 2^{-k \beta +1}B''_k) \mathrm{diag}(\nu'') + W_{k}\label{eq:splitB''}
\end{align}
with double-precision shift values $\mu'' \in \mathbb{F}_{64}^m$, $\nu'' \in \mathbb{F}^p$ and 8-bit integer matrices $A''_i \in \mathbb{I}_{8}^{m \times n}$, $B''_i\in \mathbb{I}_{8}^{n \times p}$ for $i=1,\dots,k$.

\begin{algorithm}[htb]
	\caption{Mixed-precision splitting method for Ozaki scheme via bit masking from~\cite{ootomo2024dgemm}. This algorithm is specialized for the Ozaki scheme using the INT8 Tensor Core.\label{alg:OzakiSplit-ootomo}}
	\begin{algorithmic}[1]
		\REQUIRE $A \in \mathbb{F}_{64}^{m \times n}$, $k \in \mathbb{N}$
		\ENSURE $A''_s \in \mathbb{I}_8^{m \times n}$, $\mu'' \in \mathbb{F}_{64}^m$, $s=1,\dots,k$
		\STATE $\beta \gets \min( 7, \lfloor (31 - \log_2 n)/2 \rfloor )$
		\STATE $\mu''_i \gets 2^{\lfloor \log_2 \max_j |a_{ij}| \rfloor}$ for $i=1,\dots,m$
		\FOR{$s=1,\dots,k$}
		\STATE Extract $s$th $\beta$ bits of mantissa of $a_{ij}$ $\forall i,j$ via bit masking and hold those as INT8 variable $(A''_s)_{ij}$
		\ENDFOR
	\end{algorithmic}
\end{algorithm}

After splitting matrices as in \eqref{eq:splitA''} and \eqref{eq:splitB''}, 
\begin{align*}
AB 
&= \sum_{s+t \le k+1} \mathrm{diag}(\mu'') 2^{-\beta s+1} 2^{-\beta t+1} A''_s B''_t \mathrm{diag}(\nu'')\\
&+ \sum_{s+t > k+1} \mathrm{diag}(\mu'') 2^{-\beta s+1} 2^{-\beta t+1} A''_s B''_t \mathrm{diag}(\nu'')\\
&+ \sum_{s=1}^k \mathrm{diag}(\mu'') 2^{-\beta s+1} A''_s W_{k}\\
&+ \sum_{s=1}^k V_{k} 2^{-\beta s+1} B''_s \mathrm{diag}(\nu'')\\
&+ V_{k}W_{k}
\end{align*}
holds and there is no error in $A''_sB''_t$ for $1 \le s,t \le k$ on the INT8 Tensor Core barring overflow.
In~\cite{ootomo2024dgemm}, the approximation of $AB$ is computed using only matrix multiplications with the INT8 Tensor Core, as 
\begin{equation}\label{eq:ootomo_sum}
	AB \approx \sum_{s+t \le k+1} \mathrm{diag}(\mu'') 2^{-\beta s+1} 2^{-\beta t+1} A''_s B''_t \mathrm{diag}(\nu''),
\end{equation}
shown in Algorithm~\ref{alg:Ozakimul-ootomo}.
For larger $k$, $|(W_{k})_{ij}|$ and $|(V_k)_{ij}|$ are smaller and the approximation is more accurate.

\begin{algorithm}[htb]
	\caption{Mixed-precision matrix multiplication method for Ozaki scheme using INT8 Tensor Core from~\cite{ootomo2024dgemm}\label{alg:Ozakimul-ootomo}}
	\begin{algorithmic}[1]
		\REQUIRE $A''_s \in \mathbb{I}_8^{m \times n}$, $\mu''^{(s)} \in \mathbb{F}_{64}^m$, $B''_s \in \mathbb{I}_8^{n \times p}$, $\nu''^{(s)} \in \mathbb{F}_{64}^p$, $s=1,\dots,k$
		\ENSURE $C \in \mathbb{F}_{64}^{m \times p}$
		\STATE $\beta \gets \min( 7, \lfloor (31 - \log_2 n)/2 \rfloor )$
		\STATE $C \gets O_{m,p}$
  \hfill\COMMENT{$C'' \in \mathbb{I}_{32}^{m \times p}$}
		\FOR{$g = 2,\dots,k+1$}
		\FOR{$s = 1,\dots,g-1$}
		\STATE \label{alg:Ozakimul-ootomo-mul}$C'' \gets A''_sB''_{g-s}$
		\hfill\COMMENT{Compute using \texttt{GEMM} on INT8 Tensor Core}
		\STATE \label{alg:Ozakimul-ootomo-accum}$C \gets C +$\\
            $\mathrm{diag}(\mu'') 2^{-\beta s+1}2^{-\beta (g-s)+1} \mathrm{FP64}(C'') \mathrm{diag}(\nu'')$
		\hfill\COMMENT{Compute in FP64}
		\ENDFOR
		\ENDFOR
	\end{algorithmic}
\end{algorithm}

\begin{figure}[htb]
	\centering
		\centering
		\includegraphics[width=.6\hsize]{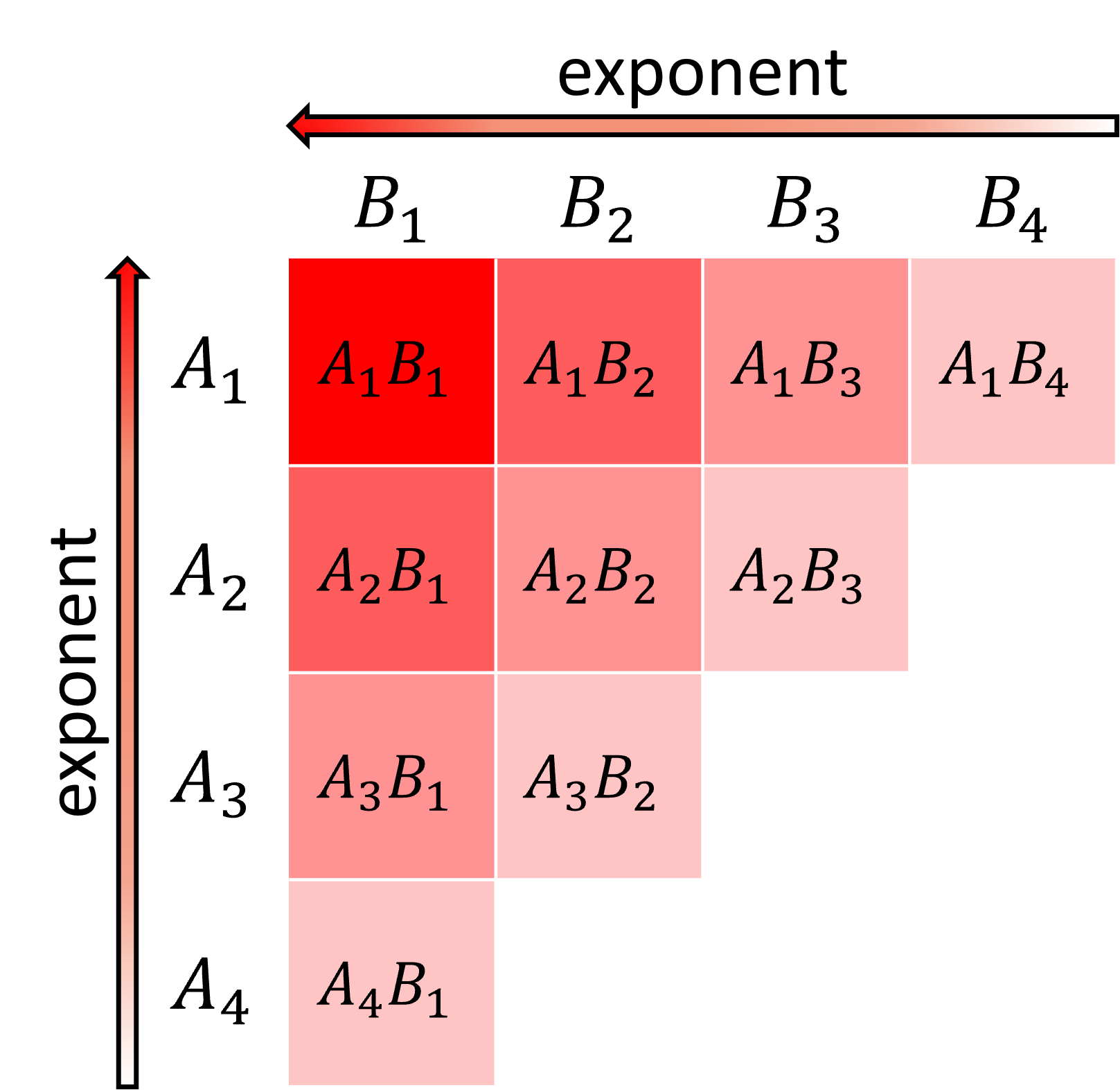}
	\caption{Images of matrix multiplications in Ozaki scheme for $k=4$}
	\label{fig:image_multiplication}
\end{figure}

Figure~\ref{fig:image_multiplication} represents images of Algorithms~\ref{alg:Ozakimul-mukunoki} and \ref{alg:Ozakimul-ootomo}.
In Figure~\ref{fig:image_multiplication}, $A_i := \mathrm{diag}(\mu'^{(i)}) A'_i$ and $B_i := B'_i \mathrm{diag}(\nu'^{(i)})$ for Algorithms~\ref{alg:OzakiSplit-mukunoki} and \ref{alg:Ozakimul-mukunoki}, and $A_i := \mathrm{diag}(\mu'')2^{-i \beta +1}A''_{i}$ and $B_i := 2^{-i \beta +1}B''_{i}\mathrm{diag}(\nu'')$ for Algorithms~\ref{alg:OzakiSplit-ootomo} and \ref{alg:Ozakimul-ootomo}.
The exponent of $A_iB_j$ is larger for smaller indices of $i,j$, so those terms have stronger effects on the accuracy of the final result.

\section{Proposed methods}
\label{sec:Proposed methods}

\subsection{Splitting with rounding to nearest}
\label{subsec:Proposed1}

We begin by explaining our motivation with a simple example.
Algorithm~\ref{alg:OzakiSplit-ootomo} described in Section~\ref{sec:Previous study} used bitmask for the splitting.
We assume that each number keeps three bits.
Let a floating-point number be $x := 377$, represented by $(101011111)_2$ in binary format.
The bitmask strategy divides $x$ into three numbers $x = x_1 + x_2 + x_3$:
\begin{equation}\label{eq:bitexample01}\begin{cases}
x_1 = (\underline{101}000000)_2,\\
x_2 = (000\underline{011}000)_2,\\
x_3 = (000000\underline{111})_2.
\end{cases}\end{equation}
In decimal format, 
\[
x_1 = 256+0+64, \quad x_2 = 0+16+8, \quad x_3 = 4+2+1.
\]
Here $x_1$ is not the nearest number to $x$.

On the other hand, let $x_1$ and $x_2$ be the nearest number to $x$ and $x-x_1$, respectively. 
Then,
\begin{equation}\label{eq:bitexample02}\begin{cases}
x_1 = (\underline{101}000000)_2,\\
x_2 = (000\underline{100}000)_2,\\
x_3 = -(00000000\underline{1})_2.
\end{cases}\end{equation}
In decimal format, 
\[
x_1 = 256+0+64, \quad x_2 = 32, \quad x_3 = -1.
\]
There are two drawbacks in the bitmask strategy.
One drawback is that even if the leading bit is zero, we keep it, so that the number of significant figures decreases as in $x_2$ in~\eqref{eq:bitexample01}.
The other drawback is that the truncation error can be reduced.
For example, if we set two slices, there is a difference in the truncation error $|x_3|$ in~\eqref{eq:bitexample01} and~\eqref{eq:bitexample02}.
It is expected that the nearest strategy makes the truncation error small.
Summarizing, even if we obtain a computed result with $k$ slices using the rounding to nearest strategy, the accuracy may be comparable to that with $k+1$ slices using the bitmask strategy described in Section~\ref{sec:Previous study} as the previous study.
In this case, we must carefully monitor the following via numerical examples:
\begin{itemize}
	\item the increase in the splitting cost because the cost for the bitmask is low;
	\item the reduction in the number of matrix multiplications; and
	\item the reduction in the term for the accumulation.
\end{itemize}
Note that in the worst case, the results of the splitting with bitmask and the nearest strategy are the same.
It can be explained using the following example: 
\[
x := (100010001)_2.
\]
For the bit mask strategy, we have
\[\begin{cases}
    x_1 = (\underline{100}000000)_2,\\
    x_2 = (000\underline{010}000)_2,\\
    x_3 = (000000\underline{001})_2.
\end{cases}
\]
For the rounding to the nearest strategy, we obtain
\[\begin{cases}
    x_1 = (\underline{100}000000)_2,\\
    x_2 = (0000\underline{100}00)_2,\\
    x_3 = (00000000\underline{1})_2.
\end{cases}
\]

The following is an algorithm for the splitting with the round to nearest strategy.

\begin{algorithm}[htb]
	\caption{Proposed method for mixed-precision splitting for Ozaki scheme between FP64 and INT8 using floating-point arithmetic in round-to-nearest-even mode with Minamihata's technique from~\cite{minamihata2016improved}\label{alg:OzakiSplit-proposal-1}}
	\begin{algorithmic}[1]
		\REQUIRE $A \in \mathbb{F}_{64}^{m \times n}$, $k \in \mathbb{N}$
		\ENSURE $A'_s \in \mathbb{I}_{8}^{m \times n}$, $\mu'^{(s)} \in \mathbb{F}_{64}^m$, $s=1,\dots,k$
		\STATE $\beta \gets \min( 7, \lfloor (31 - \log_2 n)/2 \rfloor )$
		\FOR{$s=1,\dots,k$}
		\STATE\label{alg:OzakiSplit-proposal-1_mu} $\mu'^{(s)}_i \gets 2^{\lceil \log_2 \max_j |a_{ij}| \rceil} \cdot 2^{1-\beta}$ for $i=1,\dots,m$ 
		\STATE\label{alg:OzakiSplit-proposal-1_sigma} $\sigma_i = 0.75 \cdot 2^{53} \cdot \mu'^{(s)}_i$ for $i=1,\dots,m$
		\STATE $(A_s)_{ij} = (a_{ij} + \sigma_i) - \sigma_i$ for $i=1,\dots,m$
		\hfill\COMMENT{Extract in round-to-nearest-even mode}
		\STATE $(A'_s)_{ij} \gets \mathrm{INT8}((\mu'^{(s)}_i)^{-1}\cdot (A_s)_{ij})$  $\forall i,j$
		\hfill
		\COMMENT{Convert to INT8}
		\STATE $a_{ij} \gets a_{ij}-(A_s)_{ij}$  $\forall i,j$
		\ENDFOR
	\end{algorithmic}
\end{algorithm}

\subsection{Group-wise error-free accumulation}
\label{subsec:Proposed2}

Next, we propose a method for accelerating the accumulation in FP64 for ozIMMU.
Algorithm~\ref{alg:Ozakimul-ootomo} requires the computation of a sum of $k(k+1)/2$ FP64 matrices at the \ref{alg:Ozakimul-ootomo-accum}th line.
The accumulation accounts for a large ratio of the total computation time of ozIMMU as shown in Figures~\ref{fig:time_breakdown_ozIMMU_RTX4090.png} and \ref{fig:time_breakdown_ozIMMU_gh200.png}. 
For this challenge, we propose a method for accelerating ozIMMU by reducing the number of additions in FP64.

Define $\mathbb{G}_g \subset \mathbb{R}^2$ for $g=3,\dots,k+1$ as
\[
\mathbb{G}_g := \{(i,j) \mid i+j = g\}.
\]
Recall that Algorithm~\ref{alg:Ozakimul-ootomo} performs
\begin{equation}\label{eq:accum_original}
\begin{aligned}
&C \gets \\
&\sum_{t=2}^{k+1} \sum_{s=1}^{g-1} \mathrm{diag}(\mu'') 2^{-\beta s+1}2^{-\beta (g-s)+1} A''_{s}B''_{g-s} \mathrm{diag}(\nu'').
\end{aligned}
\end{equation}
Let $(i_1,j_1),\dots,(i_{g-1},j_{g-1}) \in \mathbb{G}_g$
such that $i_s \neq i_t$ and $j_s \neq j_t$ implies $s \neq t$ for $1 \le s, t \le g-1$, where $g \in \{3,\dots,k+1\}$.
Then, the inner sum of \eqref{eq:accum_original} can be expressed as follows:
\begin{equation}\label{eq:errfreesum}
\begin{aligned}
&\sum_{s = 1}^{g-1} \mathrm{diag}(\mu'') 2^{-\beta i_s+1}2^{-\beta j_s+1} A''_{i_s}B''_{j_s} \mathrm{diag}(\nu'')\\
&= 2^{-\beta g+2} \mathrm{diag}(\mu'') \left( \sum_{s = 1}^{g-1} A''_{i_s}B''_{j_s} \right) \mathrm{diag}(\nu'')
\end{aligned}
\end{equation}
which uses that $i_s + j_s = g$ for all $s=1,\dots,g-1$.
If no overflow occurs in $A''_{i_s}B''_{j_s} + A''_{i_t}B''_{j_t}$ ($1\le s,t \le g-1$) in INT32, the summation can be performed on the accumulator of \texttt{GEMM} on the INT8 Tensor Core.
Therefore, applying the above transformation reduces the numbers of slow conversions from INT32 to FP64, scalings, and slow summations in FP64.
Let $r \in \mathbb{N}$ be 
\begin{equation}
r := \max(1,2^{31 - 2\beta  - \lceil\log_2 n \rceil}).
\label{eq:def_r}    
\end{equation}
Then, the summation
\[
\sum_{s = 1}^{\min(r,g-1)} A''_{i_s}B''_{j_s}
\]
can be computed without error using \texttt{GEMM} on the INT8 Tensor Core.
The summation of $r$ instances of $A''_{i_s}B''_{j_s}$ can be computed without error; however, the summation of all $|\mathbb{G}_g|$ instances of $A''_{i_s}B''_{j_s}$ cannot always be computed without error.
The proof validating error-free summation is provided in Section~\ref{sec:Rounding error analysis}.
Finally, we present Algorithm~\ref{alg:errfreesum}, which has a reduced number of summands in FP64 accumulation.

\begin{algorithm}[htb]
	\caption{Proposed method for groupwise error-free accumulation for Ozaki scheme using INT8 Tensor Core\label{alg:errfreesum}}
	\begin{algorithmic}[1]
		\REQUIRE $A''_s \in \mathbb{I}_8^{m \times n}$, $\mu''^{(s)} \in \mathbb{F}_{64}^m$, $B''_s \in \mathbb{I}_8^{n \times p}$, $\nu''^{(s)} \in \mathbb{F}_{64}^p$, $s=1,\dots,k$
		\ENSURE $C \in \mathbb{F}_{64}^{m \times p}$
		\STATE $\beta \gets \min( 7, \lfloor (31 - \log_2 n)/2 \rfloor )$
		\STATE $r \gets \max(1,2^{31 - 2\beta - \lceil\log_2 n \rceil})$
		\hfill \COMMENT{{\#}addends for error-free accum.}
		\STATE $C \gets O_{m,p}$
		
		\FOR{$g = 2,\dots,k+1$}
		\STATE $q \gets 0$
		\STATE $C'' \gets O_{m,p}$
  \hfill\COMMENT{$C'' \in \mathbb{I}_{32}^{m \times p}$}
		\FOR{$s = 1,\dots,g-1$}
		\STATE $q \gets q+1$
		\STATE $C'' \gets C'' + A''_sB''_{g-s}$
		\hfill\COMMENT{Compute using \texttt{GEMM} on INT8 Tensor Core}
		\IF{$q == r\ \|\ s == g-1$}
		\STATE $C \gets C + 2^{-\beta g+2} \mathrm{diag}(\mu'') \mathrm{FP64}(C'') \mathrm{diag}(\nu'')$
		\hfill\COMMENT{Compute in FP64}
		\STATE $q \gets 0$
		\STATE $C'' \gets O$
		\ENDIF
		\ENDFOR
		\ENDFOR
		
	\end{algorithmic}
\end{algorithm}

Assuming $r \ge k$ for simplify, all $A''_iB''_j$ for $(i,j) \in \mathbb{G}_g$ can be accumulated without overflow for $g=3,\dots,k+1$.
Algorithm~\ref{alg:errfreesum_simple} shows the Ozaki scheme with groupwise error-free accumulation for $r \ge k$.

\begin{algorithm}[htb]
	\caption{Simple version of proposed method for groupwise error-free accumulation for Ozaki scheme using INT8 Tensor Core for $r \ge k$\label{alg:errfreesum_simple}}
	\begin{algorithmic}[1]
		\REQUIRE $A''_s \in \mathbb{I}_8^{m \times n}$, $\mu''^{(s)} \in \mathbb{F}_{64}^m$, $B''_s \in \mathbb{I}_8^{n \times p}$, $\nu''^{(s)} \in \mathbb{F}_{64}^p$, $s=1,\dots,k$
		\ENSURE $C \in \mathbb{F}_{64}^{m \times p}$
		\STATE $\beta \gets \min( 7, \lfloor (31 - \log_2 n)/2 \rfloor ) = 7$
		\STATE $C \gets O_{m,p}$
		\FOR{$g = 2,\dots,k+1$}
            \STATE $C'' \gets O_{m,p}$
  \hfill\COMMENT{$C'' \in \mathbb{I}_{32}^{m \times p}$}
		\FOR{$s = 1,\dots,g-1$}
		\STATE $C'' \gets C'' + A''_sB''_{g-s}$
		\hfill\COMMENT{Compute using \texttt{GEMM} on INT8 Tensor Core}
		\ENDFOR
		\STATE $C \gets C + \mathrm{diag}(\mu'') 2^{-\beta g+2} \mathrm{FP64}(C'') \mathrm{diag}(\nu'')$
		\hfill\COMMENT{Compute in FP64}
		\ENDFOR
	\end{algorithmic}
\end{algorithm}

\subsection{Combination of proposals}
\label{subsec:Proposed3}
Next, we accelerate ozIMMU by reducing the number of summands in FP64 accumulation and improving the splitting method.
For this purpose, we combine the methods proposed in Sections~\ref{subsec:Proposed1} and \ref{subsec:Proposed2}.
In order to use group-wise error-free accumulation as in Algorithm~\ref{alg:errfreesum}, $A$ and $B$ need to be expressed as in \eqref{eq:splitA''} and \eqref{eq:splitB''}.
Thus, we determine 
\begin{align*}
	\mu''_i &\gets 2^{\lceil \log_2 \max_j |a_{ij}| \rceil} \cdot 2^{1-\beta},\\
	\sigma_i &\gets 0.75 \cdot 2^{53-\beta (s-1)}\cdot \mu''_i
\end{align*}
as the \ref{alg:OzakiSplit-proposal-1_mu}rd and \ref{alg:OzakiSplit-proposal-1_sigma}th lines in Algorithm~\ref{alg:OzakiSplit-proposal-1}. 
Then, $A_s$ are extracted using rounding to nearest floating-point arithmetic as
\[
(A_s)_{ij} \gets (a_{ij} + 2^{\beta (s-1)}\sigma_i) - 2^{\beta (s-1)}\sigma_i
\]
for the constant $2^{\beta (s-1)}\sigma_i$ with a common ratio of $2^\beta$.
Finally, we obtain Algorithm~\ref{alg:OzakiSplit-proposal-3}.
The shift values for $A_s$ are $\mu''^{(s)}_i \cdot 2^{\beta(s-1)}$; thus, $\mu''^{(s)}_i$ are determined before the ``for'' statement and the shift values for $A_s$ for $s \ge 1$ are calculated by shifting $\mu''^{(s)}_i$ by a constant ratio $2^{\beta}$.
Hence, the algorithm finds the maximum absolute values $\max_j |a_{ij}|$ for $i=1,\dots,m$ only once before the ``for'' statement.
On the other hand, the shift values $\mu'^{(s)}_i$ are determined inside the ``for'' statement in Algorithm~\ref{alg:OzakiSplit-proposal-1}.
Therefore, the algorithm finds the maximum absolute values $\max_j |a_{ij}|$ for $i=1,\dots,m$ at most $k$ times.

\begin{algorithm}[htb]
	\caption{Another proposed method for mixed-precision splitting for Ozaki scheme between FP64 and INT8 using floating-point arithmetic in round-to-nearest-even mode with Minamihata's technique from~\cite{minamihata2016improved}. The results can be expressed as in \eqref{eq:splitA''} for error-free group-wise accumulation.\label{alg:OzakiSplit-proposal-3}}
	\begin{algorithmic}[1]
		\REQUIRE $A \in \mathbb{F}_{64}^{m \times n}$, $k \in \mathbb{N}$
		\ENSURE $A''_s \in \mathbb{I}_8^{m \times n}$, $\mu'' \in \mathbb{F}_{64}^m$, $s=1,\dots,k$
		\STATE $\beta \gets \min( 7, \lfloor (31 - \log_2 n)/2 \rfloor )$
		\STATE $\mu''_i \gets 2^{\lceil \log_2 \max_j |a_{ij}| \rceil} \cdot 2^{1-\beta}$ for $i=1,\dots,m$
		\FOR{$s=1,\dots,k$}
		\STATE $\sigma_i \gets 0.75 \cdot 2^{53-\beta (s-1)}\cdot \mu''_i$ for $j=1,\dots,m$
		\STATE $(A_s)_{ij} \gets (a_{ij} + \sigma_i) - \sigma_i$  $\forall i,j$
		\hfill\COMMENT{Extract in round-to-nearest-even mode}
		\STATE $(A''_s)_{ij} \gets \mathrm{INT8}((\mu''^{(s)}_i)^{-1}\cdot 2^{-\beta(s-1)}\cdot (A_s)_{ij})\ \forall i,j$
		\COMMENT{Convert to INT8}
		\STATE $a_{ij} \gets a_{ij}-(A_s)_{ij}$  $\forall i,j$
		\ENDFOR
	\end{algorithmic}
\end{algorithm}

\section{Numerical examples}
\label{sec:Numerical examples}

We have implemented the proposed methods in~\cite{ozIMMU-uchino}, replacing specific code in Ootomo's ozIMMU library with our code.
All numerical experiments were conducted on NVIDIA GH200 Grace Hopper Superchip and NVIDIA GeForce RTX 4090 with the GNU C++ Compiler 11.4.1 and NVIDIA CUDA Toolkit 12.5.82.
The tested methods will be denoted as follows:
\begin{itemize}
	\item \texttt{ozIMMU-}$k$\texttt{\phantom{\_EF}}: Ootomo's implementation with $k$ slices
	\item \texttt{ozIMMU\_RN-}$k$: Proposed method in Section~\ref{subsec:Proposed1} with $k$ slices
	\item \texttt{ozIMMU\_EF-}$k$: Proposed method in Section~\ref{subsec:Proposed2} with $k$ slices
	\item \texttt{ozIMMU\_H-}$k$\texttt{\phantom{F}}: Proposed method in Section~\ref{subsec:Proposed3} with $k$ slices
\end{itemize}

\subsection{Accuracy}
\label{subsec:Accuracy}

\begin{figure}[htb]
	\centering
	\includegraphics[width=\hsize]{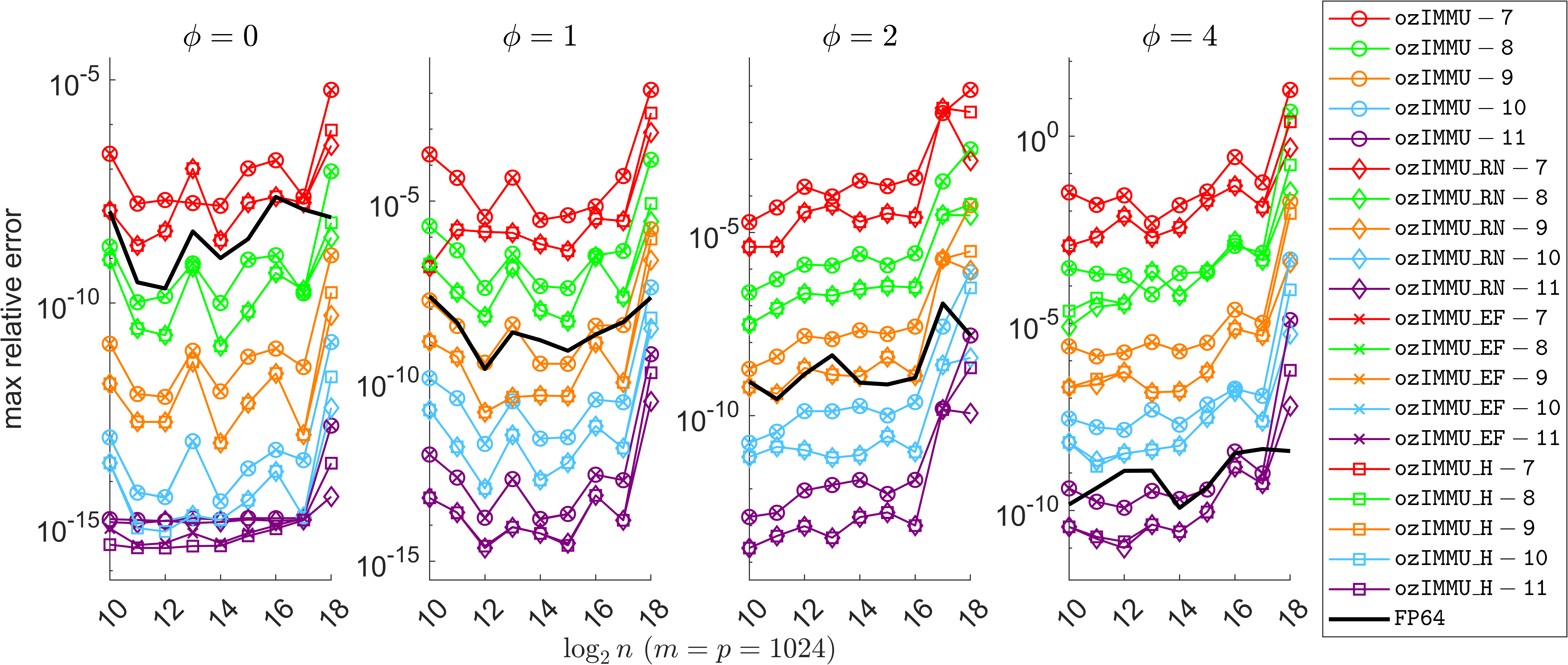}
	\caption{Comparison of accuracy between \texttt{ozIMMU} and proposed methods. Matrix $A$ has entries $a_{ij} := (U_{ij}-0.5)\cdot \exp(\phi\cdot N_{ij})$, where $U_{ij} \in (0,1)$ are uniformly distributed random numbers and $N_{ij}$ are drawn from the standard normal distribution, for $1 \le i,j \le m$ and $m=n=p$. Matrix $B$ has a similar composition.}
	\label{fig:accuracy_huge.png}
\end{figure}

Figure~\ref{fig:accuracy_huge.png} shows the accuracies of \texttt{ozIMMU-}$k$, \texttt{ozIMMU\_RN-}$k$, \texttt{ozIMMU\_EF-}$k$, and \texttt{ozIMMU\_H-}$k$ for $k=7,\dots,11$.
We set $A \in \mathbb{F}_{64}^{n \times n}$ as $a_{ij} := (U_{ij}-0.5)\cdot \exp(\phi\cdot N_{ij})$, where $U_{ij} \in (0,1)$ are uniformly distributed random numbers and $N_{ij}$ are drawn from the standard normal distribution, for $1 \le i,j \le n$.
The constant $\phi$ specifies the tendency of the difficulty in terms of matrix multiplication accuracy.
A matrix $B \in \mathbb{F}_{64}^{n \times n}$ is set similarly to $A$.

The accuracy deteriorated for $n > 2^{17}$ because the maximum number of significant bits $\beta = \min( 7, \lfloor (31 - \log_2 n)/2 \rfloor )$ of the elements of the split matrices is less than $7$.
The accuracies of \texttt{ozIMMU} and \texttt{ozIMMU\_EF} are comparable, as are those of, \texttt{ozIMMU\_RN} and \texttt{ozIMMU\_H}.
\texttt{ozIMMU\_RN} and \texttt{ozIMMU\_H}, which use a splitting method via rounding to nearest floating-point arithmetic, are more accurate than \texttt{ozIMMU} and \texttt{ozIMMU\_EF}, which use a splitting method via bit masking.
Occasionally, to obtain results comparable with that of FP64, \texttt{ozIMMU\_RN-}$k$, \texttt{ozIMMU\_H-}$k$, \texttt{ozIMMU-}$(k+1)$, and \texttt{ozIMMU\_EF-}$(k+1)$ are required; i.e., \texttt{ozIMMU\_RN} and \texttt{ozIMMU\_H} require fewer splits than \texttt{ozIMMU} and \texttt{ozIMMU\_EF}.
In particular, for $\phi = 2$, \texttt{ozIMMU\_RN-}$9$ and \texttt{ozIMMU\_H-}$9$ produce comparable results to FP64; however, the accuracies of the results of \texttt{ozIMMU-}$9$ and \texttt{ozIMMU\_EF-}$9$ are worse than that of FP64.
In such cases, $k=10$ is required for \texttt{ozIMMU} and \texttt{ozIMMU\_EF} to attain more accurate results than FP64.

\subsection{Time breakdown}
\label{subsec:Time breakdown}

\begin{figure}[htb]
	\centering
	\includegraphics[width=\hsize]{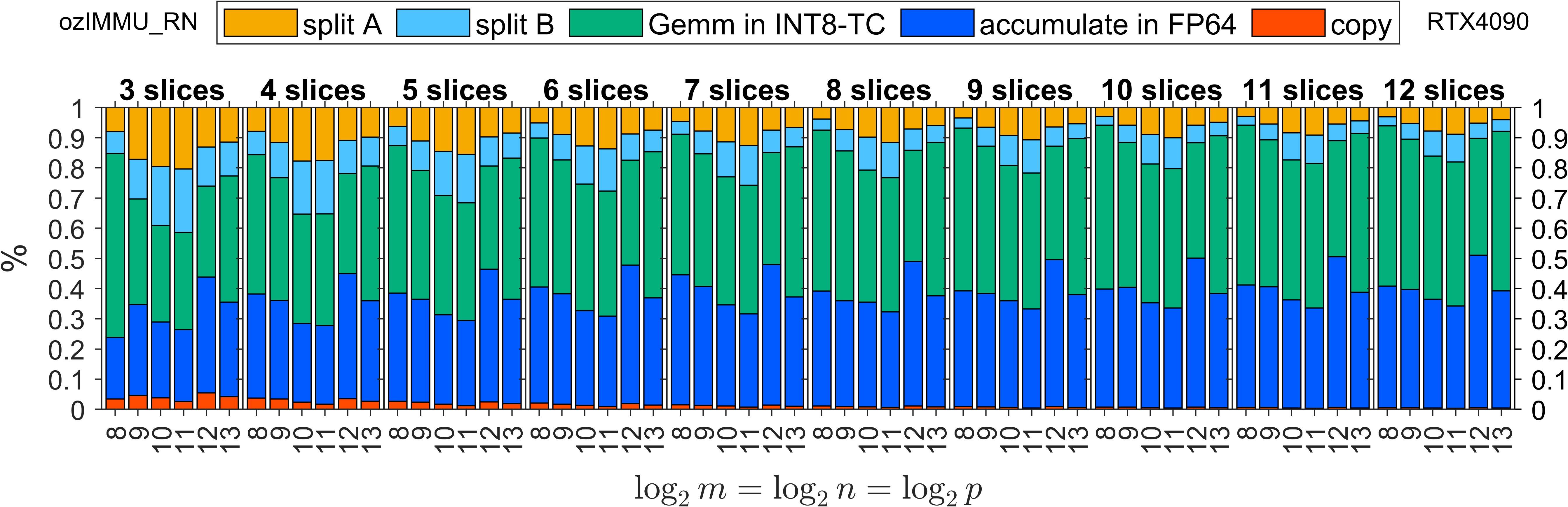}
	\caption{Time breakdown of \texttt{ozIMMU\_RN} on NVIDIA GeForce RTX 4090}
	\label{fig:time_breakdown_ozIMMU_RN_RTX4090.png}
\end{figure}

\begin{figure}[htb]
	\centering
	\includegraphics[width=\hsize]{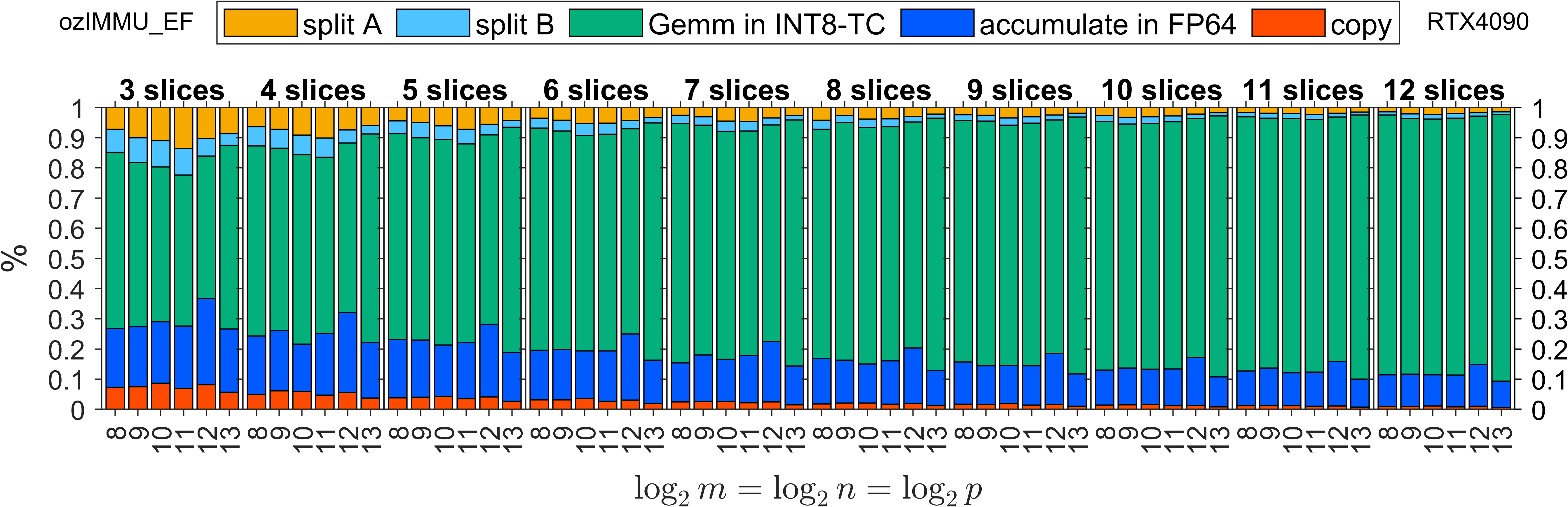}
	\caption{Time breakdown of \texttt{ozIMMU\_EF} on NVIDIA GeForce RTX 4090}
	\label{fig:time_breakdown_ozIMMU_EF_RTX4090.png}
\end{figure}

\begin{figure}[htb]
	\centering
	\includegraphics[width=\hsize]{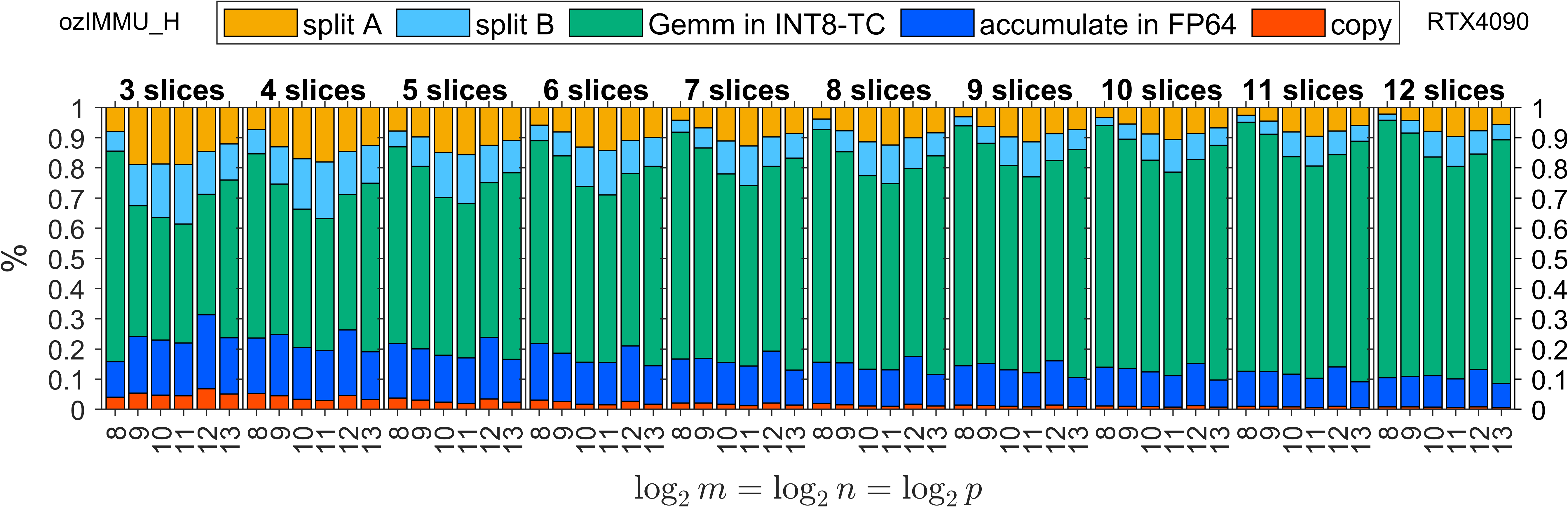}
	\caption{Time breakdown of \texttt{ozIMMU\_H} on NVIDIA GeForce RTX 4090}
	\label{fig:time_breakdown_ozIMMU_H_RTX4090.png}
\end{figure}

\begin{figure}[htb]
	\centering
	\includegraphics[width=\hsize]{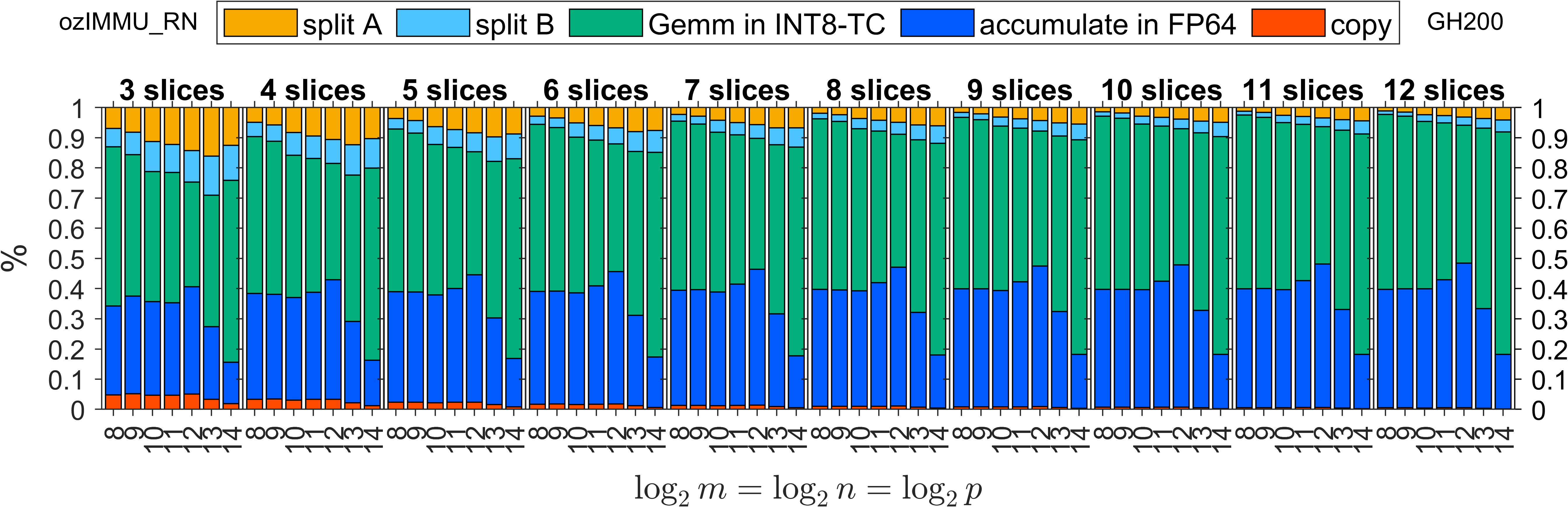}
	\caption{Time breakdown of \texttt{ozIMMU\_RN} on NVIDIA GH200 Grace Hopper Superchip}
	\label{fig:time_breakdown_ozIMMU_RN_gh200.png}
\end{figure}

\begin{figure}[htb]
	\centering
	\includegraphics[width=\hsize]{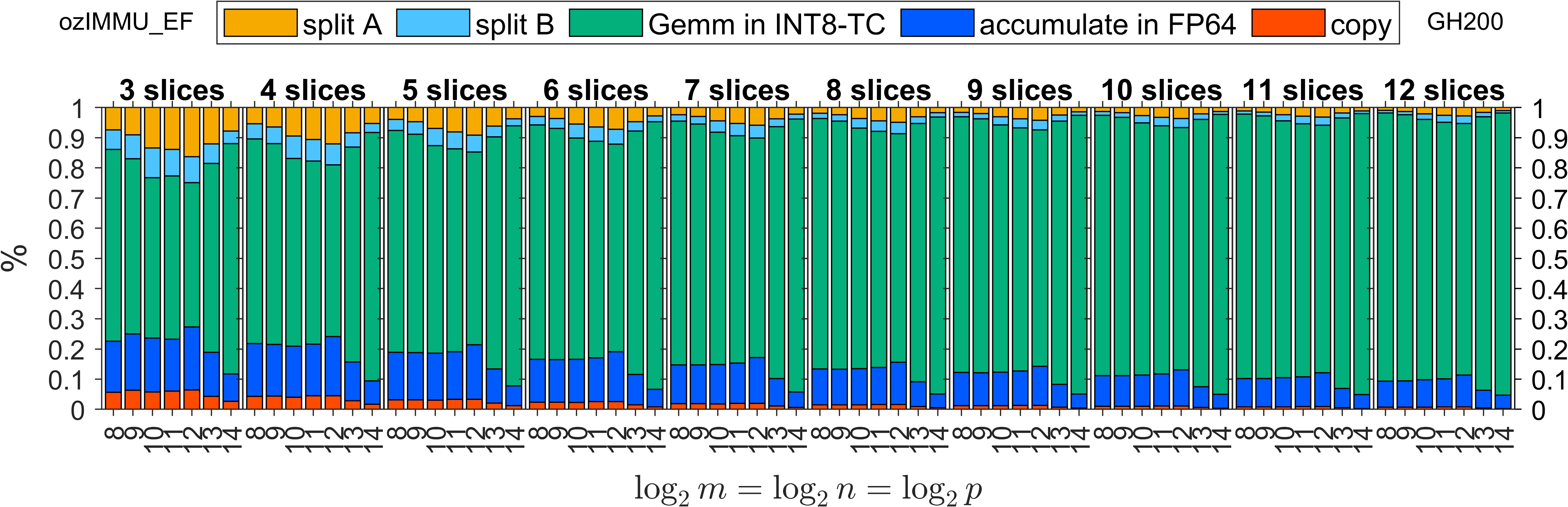}
	\caption{Time breakdown of \texttt{ozIMMU\_EF} on NVIDIA GH200 Grace Hopper Superchip}
	\label{fig:time_breakdown_ozIMMU_EF_gh200.png}
\end{figure}

\begin{figure}[htb]
	\centering
	\includegraphics[width=\hsize]{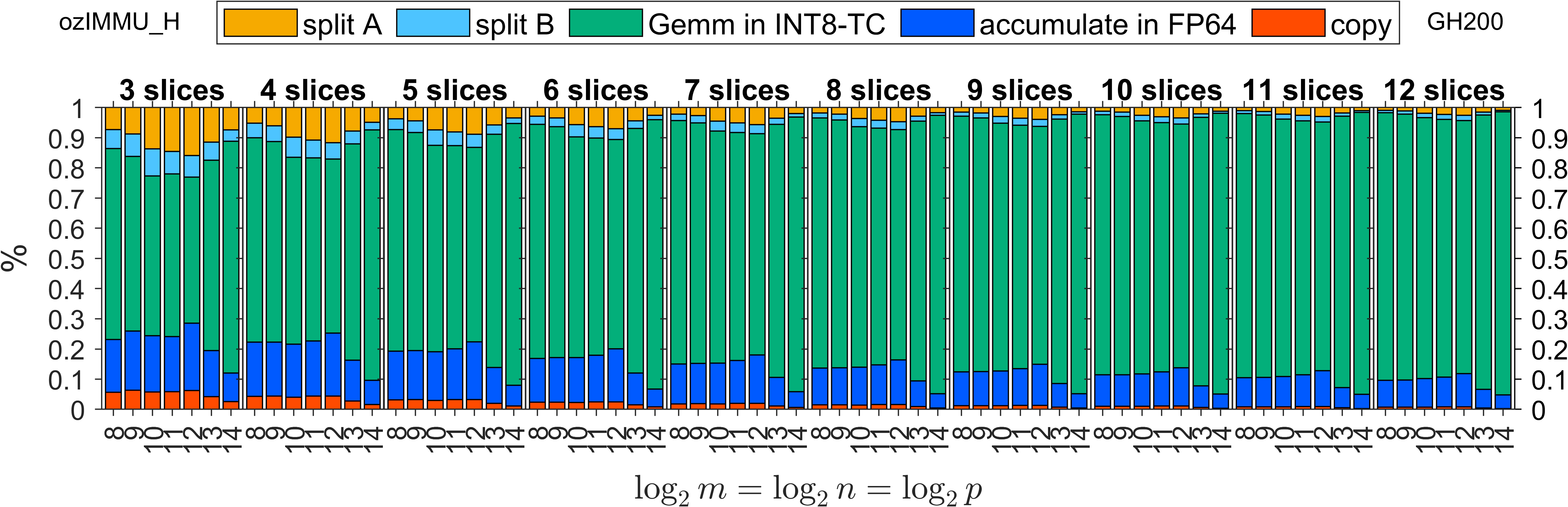}
	\caption{Time breakdown of \texttt{ozIMMU\_H} on NVIDIA GH200 Grace Hopper Superchip}
	\label{fig:time_breakdown_ozIMMU_H_gh200.png}
\end{figure}

Figures~\ref{fig:time_breakdown_ozIMMU_RN_RTX4090.png}--\ref{fig:time_breakdown_ozIMMU_H_RTX4090.png} and Figures~\ref{fig:time_breakdown_ozIMMU_RN_gh200.png}--\ref{fig:time_breakdown_ozIMMU_H_gh200.png} show time breakdowns of the proposed methods on RTX 4090 and GH200, respectively.
Note that ``split A'', ``split B'', ``Gemm in INT8-TC'', ``accumulation in FP64'', and ``copy'' correspond to the steps \ref{item1}, \ref{item2}, \ref{item3}, \ref{item4}, and \ref{item5} in the Ozaki scheme for emulating the GEMM routine as shown in Section~\ref{sec:Introduction}.

The execution time of FP64 accumulation in \texttt{ozIMMU\_RN} is not negligible, because the number of additions in FP64 is not reduced. 
The execution times of FP64 accumulation in \texttt{ozIMMU\_EF} and \texttt{ozIMMU\_H} are less than those in \texttt{ozIMMU} and \texttt{ozIMMU\_RN}.
From Figures~\ref{fig:time_breakdown_ozIMMU_RN_RTX4090.png}--\ref{fig:time_breakdown_ozIMMU_H_RTX4090.png}, the computation time of splitting via rounding to nearest floating-point arithmetic is not so fast because FP64 is much slower than the lower-precision arithmetic on RTX 4090.
From Figures~\ref{fig:time_breakdown_ozIMMU_EF_gh200.png} and \ref{fig:time_breakdown_ozIMMU_H_gh200.png}, the ratios of the computation time of splitting in \texttt{ozIMMU\_EF} and \texttt{ozIMMU\_H} are comparable on GH200.

\subsection{Performance}
\label{subsec:Performance}

\begin{figure}[htb]
	\centering
	\includegraphics[width=\hsize]{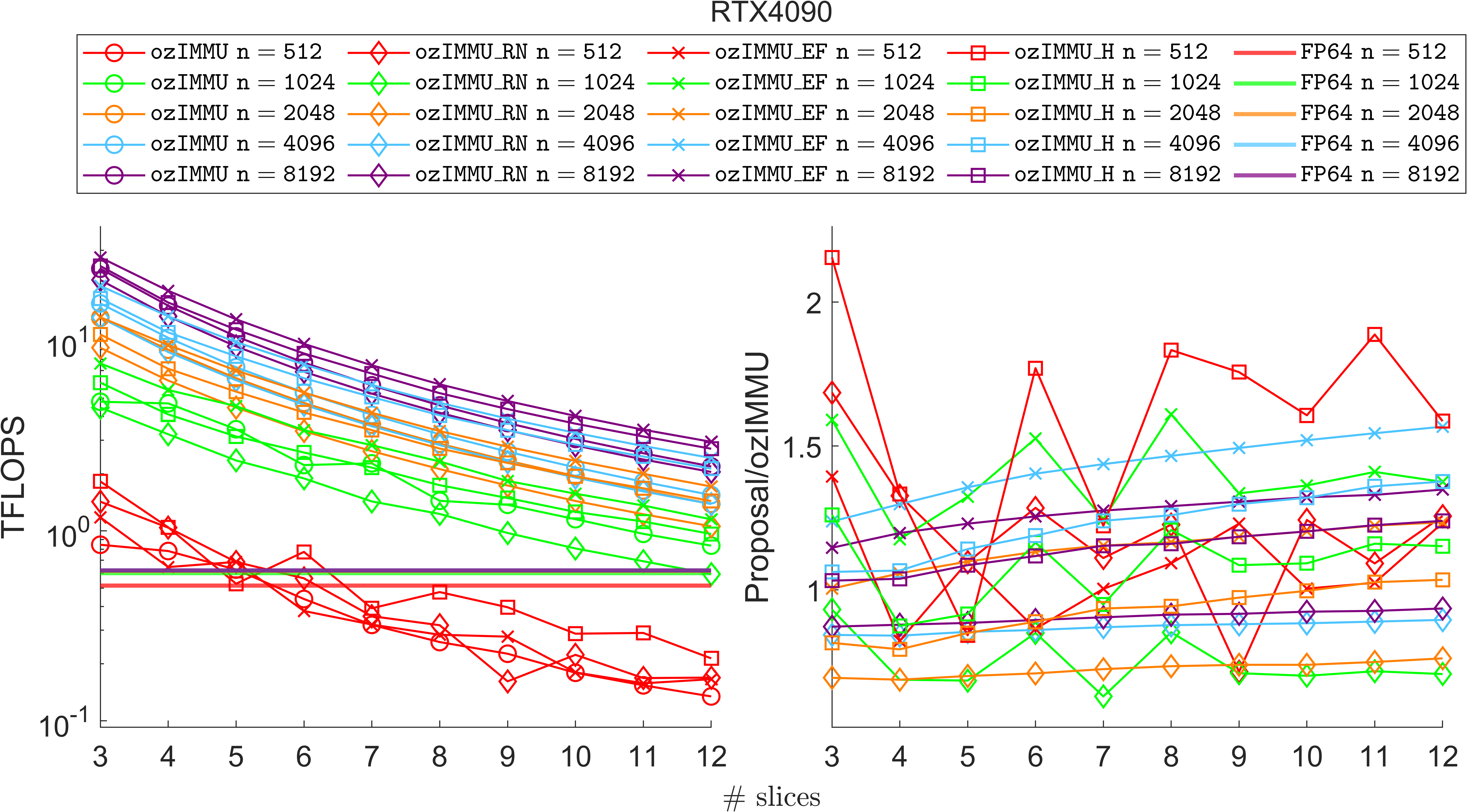}
	\caption{Throughput comparison on NVIDIA GeForce RTX 4090}
	\label{fig:flops_RTX4090.png}
\end{figure}

\begin{figure}[htb]
	\centering
	\includegraphics[width=\hsize]{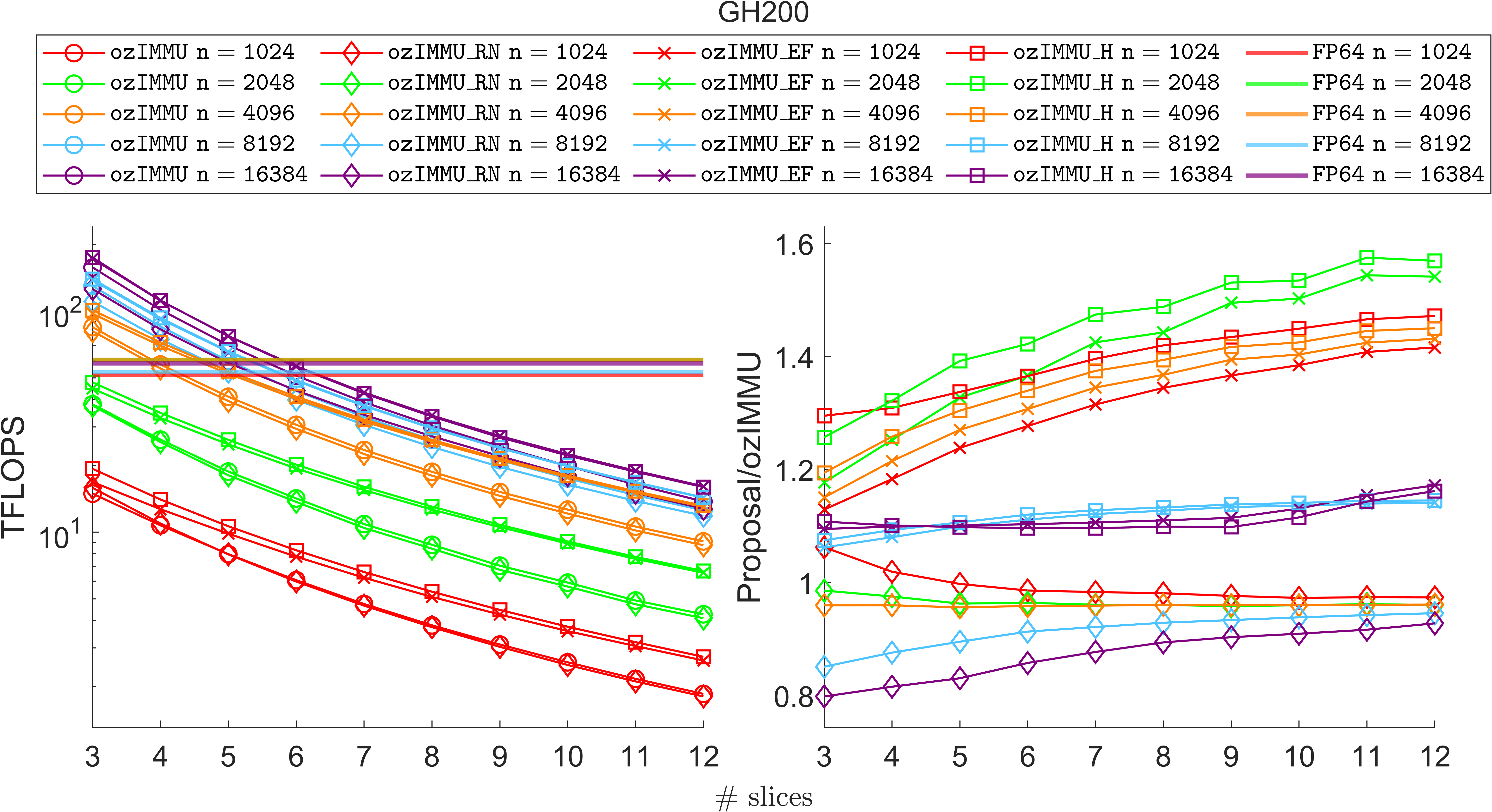}
	\caption{Throughput comparison on NVIDIA GH200 Grace Hopper Superchip}
	\label{fig:flops_gh200.png}
\end{figure}

Figures~\ref{fig:flops_RTX4090.png} and \ref{fig:flops_gh200.png} show throughput in TFLOPS and ratio to \texttt{ozIMMU}.
A smaller number of slices corresponds to better performance because the total computation cost is less.

On RTX 4090, all methods are faster than or comparable to matrix multiplication in FP64 for $n \ge 1024$ because FP64 is much slower than the lower-precision arithmetic.
\texttt{ozIMMU\_EF} and \texttt{ozIMMU\_H} are faster than \texttt{ozIMMU} almost everywhere.
In particular, \texttt{ozIMMU\_EF-}$12$ and \texttt{ozIMMU\_H-}$12$ are respectively 1.6 and 1.4 times faster than \texttt{ozIMMU} for $n = 4096$.
\texttt{ozIMMU\_RN} is slower than \texttt{ozIMMU}.
For $k=3$, \texttt{ozIMMU}, \texttt{ozIMMU\_RN}, \texttt{ozIMMU\_EF}, and \texttt{ozIMMU\_H} are $38.8$, $33.8$, $44.4$, and $40.0$ times faster than matrix multiplication in FP64 for $n = 16384$, respectively.
For $k=12$, \texttt{ozIMMU}, \texttt{ozIMMU\_RN}, \texttt{ozIMMU\_EF}, and \texttt{ozIMMU\_H} are $1.4$, $1.0$, $1.9$, and $1.6$ times faster than matrix multiplication in FP64 for $n = 1024$, respectively.
In addition, \texttt{ozIMMU}, \texttt{ozIMMU\_RN}, \texttt{ozIMMU\_EF}, and \texttt{ozIMMU\_H} are $9.4$, $8.5$, $12.0$, and $10.9$ times faster than FP64 for $k=7$ and $n = 16384$, and $7.4$, $6.8$, $9.5$, and $8.6$ times faster than FP64 for $k=8$ and $n = 16384$.

On GH200, the methods with small $k$ are faster than matrix multiplication in FP64 for $n \ge 4096$; however, the methods are much slower than FP64 otherwise.
\texttt{ozIMMU\_EF} and \texttt{ozIMMU\_H} are faster than \texttt{ozIMMU}.
In particular, \texttt{ozIMMU\_EF-}$11$ and \texttt{ozIMMU\_H-}$11$ are respectively 1.5 and 1.6 times faster than \texttt{ozIMMU} for $n = 2048$.
\texttt{ozIMMU\_RN} is not as slow on GH200 as it is on RTX 4090.
For $k=3$, \texttt{ozIMMU}, \texttt{ozIMMU\_RN}, \texttt{ozIMMU\_EF}, and \texttt{ozIMMU\_H} are $2.7$, $2.2$, $3.0$, and $3.0$ times faster than matrix multiplication in FP64 for $n = 16384$, respectively.
\texttt{ozIMMU}-$5$, \texttt{ozIMMU\_RN}-$5$, \texttt{ozIMMU\_EF-}$6$, and \texttt{ozIMMU\_H-}$6$ have comparable computation times to matrix multiplication in FP64 for $n = 16384$; however, \texttt{ozIMMU}-$k$, \texttt{ozIMMU\_RN}-$k$, \texttt{ozIMMU\_EF-}$(k+1)$, and \texttt{ozIMMU\_H-}$(k+1)$ are slower than FP64 for $n < 16384$ or $k > 5$.

\subsection{Performance vs. Accuracy}
\label{subsec:Performance vs. Accuracy}

\begin{figure}[htb]
	\centering
	\noindent\begin{minipage}[t]{0.49\hsize}
		\centering
		\includegraphics[width=\hsize]{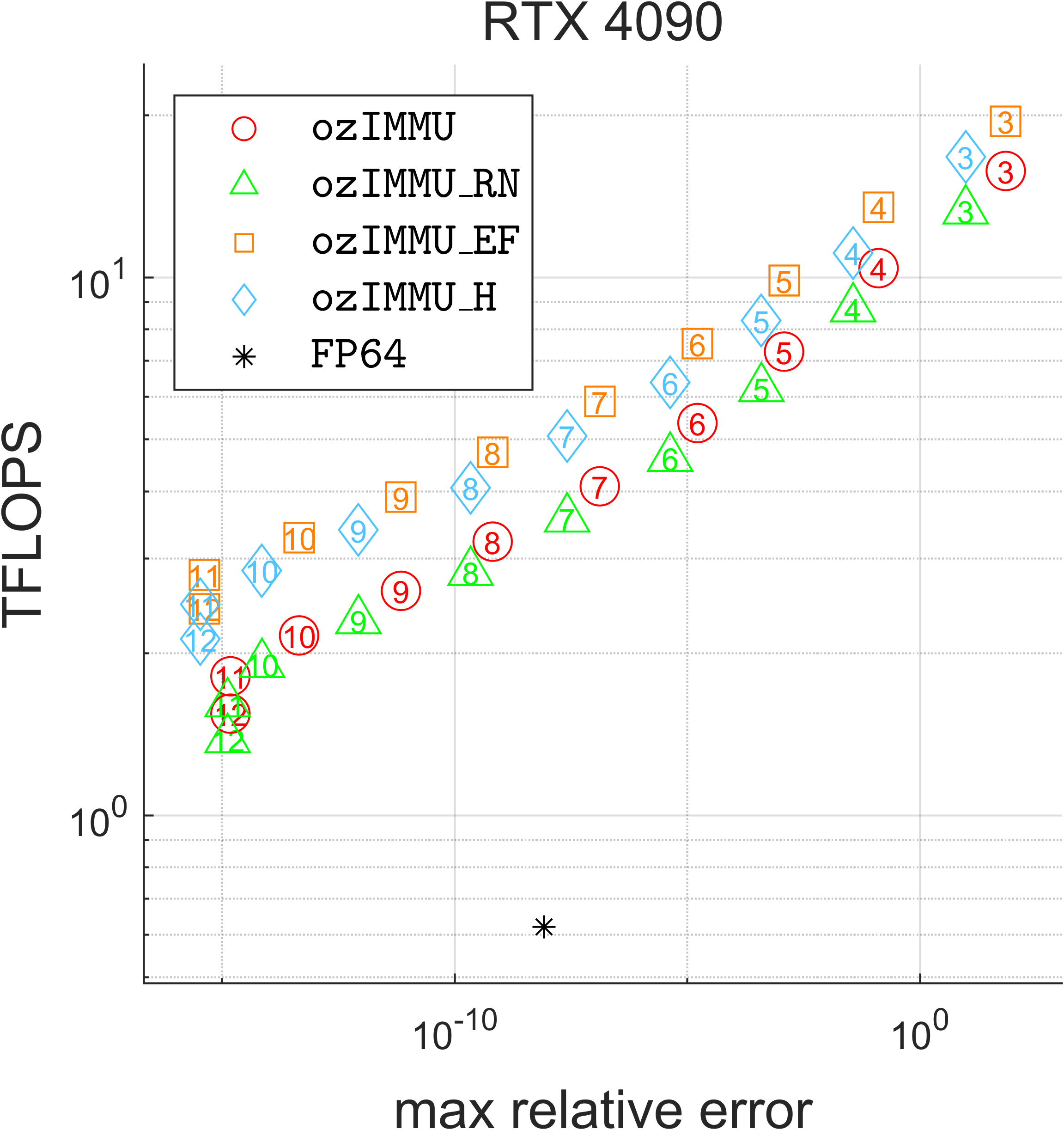}
		\subcaption{NVIDIA GeForce RTX 4090\label{subfig:flops_vs_acc_A}}
	\end{minipage}
	\hfill
	\begin{minipage}[t]{0.49\hsize}
		\centering
		\includegraphics[width=\hsize]{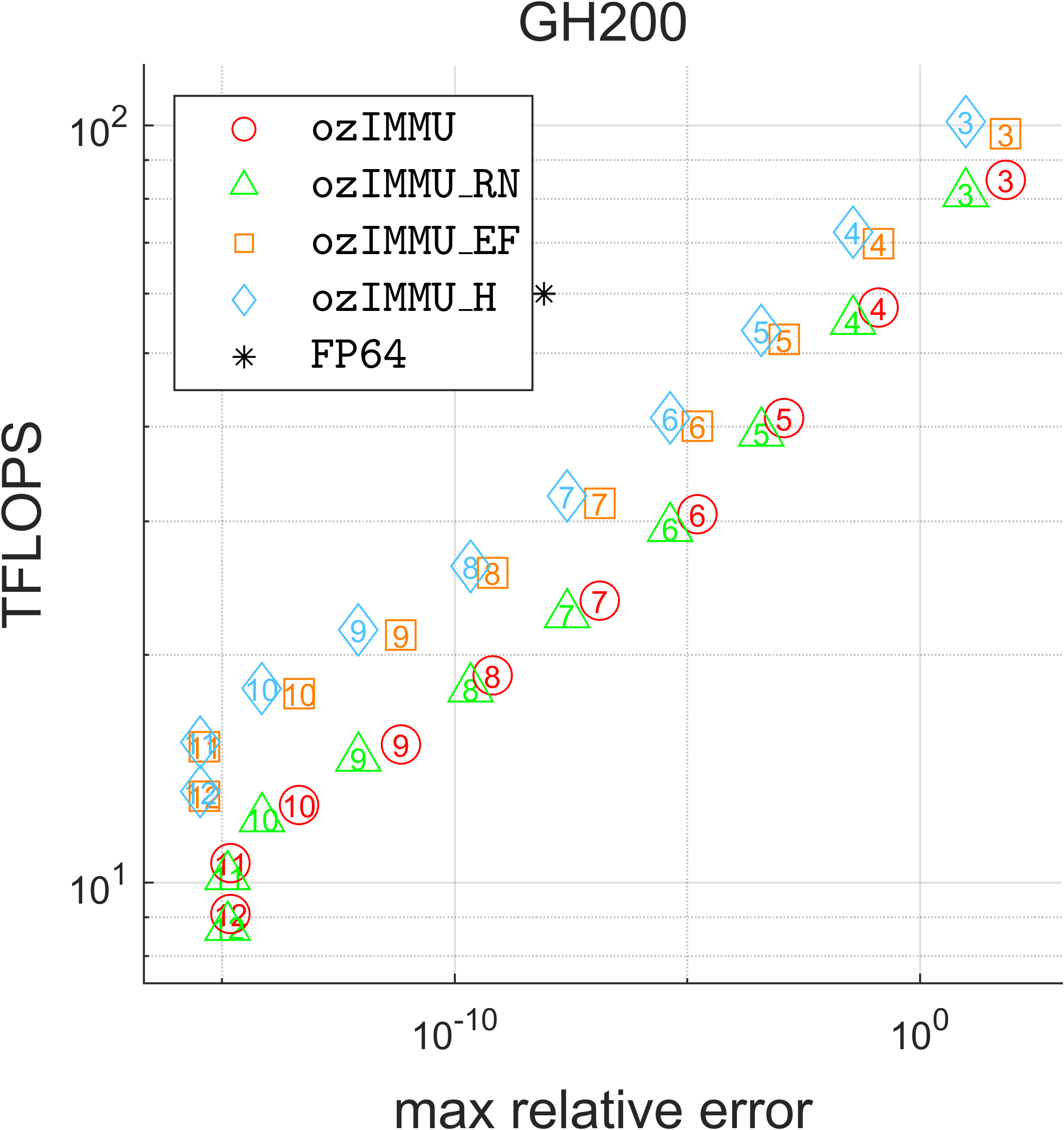}
		\subcaption{NVIDIA GH200 Grace Hopper Superchip\label{subfig:flops_vs_acc_B}}
	\end{minipage}
	\caption{Relationship between performance and accuracy for $n=4096$ and $\phi=0$}
	\label{fig:flops_vs_acc}
\end{figure}

Figure~\ref{fig:flops_vs_acc} shows throughput in a TFLOPS vs. Accuracy plot, illustrating relationships between performance and accuracy for the Ozaki scheme and for matrix multiplication in FP64 for $n=4096$ and $\phi=0$.
The numbers inside the symbols are the numbers of splices for the Ozaki scheme.

On RTX 4090, all methods are better than matrix multiplication in FP64 with respect to both performance and accuracy for $k \ge 8$.
\texttt{ozIMMU\_H} and \texttt{ozIMMU\_EF} produce results with comparable performances that are more accurate than those of \texttt{ozIMMU}.
In particular, \texttt{ozIMMU\_H-}$8$ and \texttt{ozIMMU\_EF-}$9$ produce results with comparable performances that are more accurate than those of \texttt{ozIMMU-}$7$.
On GH200 as well, \texttt{ozIMMU\_H} and \texttt{ozIMMU\_EF} produce results with comparable performances that are more accurate than those of \texttt{ozIMMU}.
In particular, \texttt{ozIMMU\_H-}$7$ and \texttt{ozIMMU\_EF-}$7$ produce results with comparable performances that are more accurate than those of \texttt{ozIMMU-}$6$.
These results also indicate that \texttt{ozIMMU\_H} and \texttt{ozIMMU\_EF} can be computed with less memory consumption than \texttt{ozIMMU}.

\section{Rounding error analysis}
\label{sec:Rounding error analysis}

In this section, we give a rounding analysis for a fixed number of slices.
We have described Algorithms~\ref{alg:OzakiSplit-ootomo} and \ref{alg:Ozakimul-ootomo} (ozIMMU), and proposed 
Algorithms~\ref{alg:OzakiSplit-proposal-1} and \ref{alg:errfreesum}, 
Algorithms~\ref{alg:OzakiSplit-ootomo} and \ref{alg:errfreesum}, and
Algorithms~\ref{alg:OzakiSplit-proposal-3} and~\ref{alg:errfreesum}.
Even if we used the rounding to nearest strategy as in Algorithms~\ref{alg:OzakiSplit-proposal-1} and~\ref{alg:OzakiSplit-proposal-3}, their error bounds would be the same as that of the bitmask strategy used in Algorithm~\ref{alg:OzakiSplit-proposal-1}.
Therefore, we focus on Algorithms~\ref{alg:OzakiSplit-ootomo} and \ref{alg:Ozakimul-ootomo} (ozIMMU) and Algorithms~\ref{alg:OzakiSplit-ootomo} and \ref{alg:errfreesum}.
Below, absolute value notation applied to a matrix means the matrix from applying absolute value element-wise.
We assume that neither overflow nor underflow occurs.
For $A \in \mathbb{F}^{m \times n}$ and $B \in \mathbb{F}^{n \times p}$, the following deterministic error bound is given by~\cite{Jeannerod2013}:
\[
|AB - \mathtt{fl}(AB)| \le nu |A| |B|.
\]
The following alternative probabilistic error bound comes from~\cite{Connolly2020} (which has the details on the assumptions and probabilities):
\[
|AB - \mathtt{fl}(AB)| \lesssim \sqrt{n}u |A| |B|.
\]
In this section, our aim is to derive an error bound on the computed result.
Let $T \in \mathbb{F}^{m \times p}$ be a computed result using $k$ slices, such as
\[
A \approx A_1 + A_2 + \dots + A_k, \quad
B \approx B_1 + B_2 + \dots + B_k.
\]
Note that for $i=1,2,\dots,k$, 
\[
A_i := \mathrm{diag}(\mu'^{(i)}) A'_i,\quad 
B_i := B'_i \mathrm{diag}(\nu'^{(i)})
\]
for Algorithms~\ref{alg:OzakiSplit-mukunoki}, \ref{alg:Ozakimul-mukunoki}, and \ref{alg:OzakiSplit-proposal-1}, and 
\[
A_i := \mathrm{diag}(\mu'')2^{-i \beta +1}A''_{i},\quad
B_i := 2^{-i \beta +1}B''_{i}\mathrm{diag}(\nu'')
\]
for Algorithms~\ref{alg:OzakiSplit-ootomo}, \ref{alg:Ozakimul-ootomo}, \ref{alg:errfreesum}, \ref{alg:errfreesum_simple}, and \ref{alg:OzakiSplit-proposal-3}.
Let $T_k$ be an approximation of $AB$ with $k$ slices.
Our aim is to obtain the upper bound on $|AB - T|$ as follows:
\begin{equation}\label{eq:basis}
\begin{aligned}
&|AB - T_k| \\
&\le \left| AB - \sum_{i=1}^k \sum_{j=1}^{k-i+1} A_i B_j \right| + \left| \sum_{i=1}^k \sum_{j=1}^{k-i+1} A_i B_j - T_k \right|.
\end{aligned}
\end{equation}
Note that we can obtain the exact product $A_i B_j$ in the above by using GEMM in the INT8 Tensor Core and scaling by powers of two.
The first term of the bound in \eqref{eq:basis} indicates the truncation error, and the second term shows an error arising in the accumulation process.
As matrix scaling does not affect rounding errors, a discussion on it is omitted.

In the following subsections, we use an upper bound provided in~\cite{Jeannerod2013} for a sum:
\begin{equation}
	\left| \sum_{i=1}^n p - \mathtt{fl}\left(\sum_{i=1}^n p \right) \right| \le (n-1)u \sum_{i=1}^n |p_i|, \quad p \in \mathbb{F}^n.
	\label{eq:error_sum}
\end{equation}
Let $\mathtt{ufp}(c)$ for $c \in \mathbb{R}$ be defined as
\[
\mathtt{ufp}(c) := 
\begin{cases}
	0 & \text{if}\ c=0, \\
	2^{\lfloor \log_2 |c| \rfloor} & \text{otherwise}.
\end{cases}
\]
The following condition is used to check whether a real number is a floating-point number.
For $c \in \mathbb{R}$, if $|c|$ is smaller than the maximum floating-point number and $c$ is an integral multiple of the minimum positive floating-point number, then it holds that
\begin{equation}
    c \in 2u \cdot \mathtt{ufp}(c)\mathbb{Z}\quad \Rightarrow\quad c \in \mathbb{F}.
    \label{eq:no_error_condition}    
\end{equation}

\subsection{Error bound for Algorithms~\ref{alg:OzakiSplit-ootomo} and \ref{alg:Ozakimul-ootomo} (ozIMMU)}
\label{sec:eb_ps}

\begin{figure}
	\centering
	\includegraphics[width=\hsize]{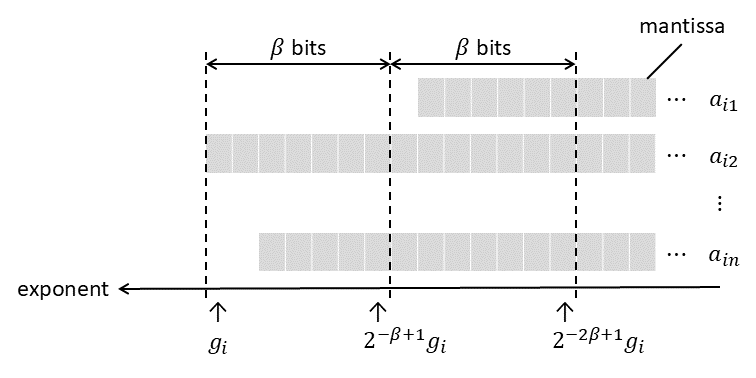}
	\caption{An image of \eqref{eq:ulp}}
	\label{fig:ulp}
\end{figure}

Let matrices $V_k \in \mathbb{F}^{m \times n}$ and $W_k \in \mathbb{F}^{n \times p}$ be defined as in \eqref{def:VW} where these show the truncation errors of $A$ and $B$ for the case of $k$ slices, respectively.
Notation $e$ indicates the vector $e=(1,1,\dots,1)^T \in \mathbb{F}^n$.
Define two vectors $g \in \mathbb{F}^m$ and $h \in \mathbb{F}^p$ related to row-wise maximums in $A$ and column-wise maximums in $B$ in the sense of the unit in the first place as
\[
g_i := \mathtt{ufp} \left( \max_j |a_{ij}| \right), \quad 
f_j := \mathtt{ufp} \left( \max_i |b_{ij}| \right).
\]
Then, from Figure~\ref{fig:ulp} and \eqref{def:VW}, we have
\begin{equation}
	|V_i| \le 2^{-\beta i+1} g e^T, \quad |W_i| \le 2^{-\beta i+1} e f^T,  
	\label{eq:VW_bound}
\end{equation}
and matrices $|A_i|$ and $|B|$ are bounded as
\begin{equation}
	|A_i| \le 2^{-\beta (i-1)+1} g e^T, \quad |B| \le 2 e f^T.
	\label{eq:each_bound}   
\end{equation}
Because 
\[
AB = (A_1 + \dots + A_k + V_k)(B_1 + \dots + B_k + W_k), 
\]
we have
\[
AB - \sum_{i=1}^k \sum_{j=1}^{k-i+1} A_i B_j = 
\sum_{i=1}^{k} A_i W_{k-i+1} + V_k B.
\]
From the last equation, \eqref{eq:VW_bound}, and~\eqref{eq:each_bound}, the truncation error is bounded as
\begin{align}
&\left| AB - \sum_{i=1}^k \sum_{j=1}^{k-i+1} A_i B_j \right| \nonumber \\
& \le 
\sum_{i=1}^{k} \left| A_i \right| \left| W_{k-i+1} \right| + \left| V_k \right| \left| B \right| \nonumber \\
& \le \sum_{i=1}^{k} 2^{-\beta (i-1)+1} ge^T \cdot 2^{-\beta (k-i+1)+1} ef^T \nonumber \\
&\quad + 2 \cdot 2^{-\beta k + 1} ge^T \cdot e f^T \nonumber \\
& = 4n \sum_{i=1}^{k} 2^{-\beta k} gf^T + 4n \cdot 2^{-\beta k} gf^T \nonumber \\
&= 4(k+1)n 2^{-\beta k} gf^T. \label{eq:truncation}
\end{align}

If $|g_i - |a_{ij}||$ for $1 \le j \le n$ and $|f_j - |b_{ij}||$ for $1 \le i \le n$ are very small, we can assume that
\begin{equation}
	4n g f^T \approx |A| |B|.
	\label{eq:well_conditioned}   
\end{equation}
In this case, we have
\begin{equation}
	\left| AB - \sum_{i=1}^k \sum_{j=1}^{k-i+1} A_i B_j \right| \lesssim (k+1) 2^{-\beta k} |A||B|.
	\label{eq:best}
\end{equation}

We next focus on an error in the accumulation, the second term in~\eqref{eq:basis}.
If $A_i$ and $B_i$ are obtained by Algorithm~\ref{alg:OzakiSplit-ootomo}, 
\begin{equation}
	|A| = \sum_{i=1}^k |A_i|, \quad |B| = \sum_{i=1}^k |B_i|
	\label{eq:abs_relation}    
\end{equation}
are satisfied.
Because we compute the sum of $\frac{1}{2}k(k+1)$ floating-point matrices, the following bound is immediately obtained from~\eqref{eq:error_sum} and~\eqref{eq:abs_relation}:
\begin{align}
&\left| \sum_{i=1}^k \sum_{j=1}^{k-i+1} A_i B_j - T_k \right|\notag\\
&\le \left( \frac{1}{2} k (k+1)-1 \right) u \sum_{i=1}^k \sum_{j=1}^{k-i+1} |A_i B_j|\notag \\
&\le \left( \frac{1}{2} k (k+1)-1 \right) u |A| |B|.
\label{eq:straightforward}
\end{align}
We will show that $k' (\le k)$ exists such that
\begin{equation}
	\sum_{s=1}^{k'} \sum_{t=1}^{k'-s+1} A_s B_t \in \mathbb{F}^{m \times n}.
	\label{eq:no_error_sum}
\end{equation}
Then, the bound~\eqref{eq:straightforward} is improved.
From Figure~\ref{fig:ulp}, we have
\begin{equation}\label{eq:ulp}
	\left( A_s \right)_{ij} \in 2^{-s\beta+1} g_i \mathbb{Z}, \quad
	\left( B_t \right)_{ij} \in 2^{-t\beta+1} f_j \mathbb{Z},
\end{equation}
and these derive
\begin{equation}
	\left( A_s \right)_{i \ell} \left( B_t \right)_{\ell j} 
	\in 2^{-\beta(s+t)+2} g_i f_j \mathbb{Z}.
	\label{eq:multiple01}
\end{equation}
Figure~\ref{fig:ulp} shows an image of \eqref{eq:ulp}.
It follows from this that 
\begin{equation}
\left( \sum_{s=1}^{k'} \sum_{t=1}^{k'-s+1} A_s B_t \right)_{ij} \in 
2^{-\beta(k'+1)+2} g_i f_j \mathbb{Z}.
\label{eq:multiple}    
\end{equation}

From Figure~\ref{fig:ulp}, the elements of the matrices are bounded as
\[
|A_s|_{ij} \le 2^{-\beta(s-1)+1} g_i, \quad
|B_t|_{ij} \le 2^{-\beta(t-1)+1} f_j,
\]
so that we have
\begin{equation}
	|A_s B_t|_{ij} \le n 2^{-\beta(s+t-2)+2} g_i f_j.
	\label{eq:upperbound}   
\end{equation}
Therefore, from~\eqref{eq:upperbound}, we have the following bound on a dot product:
\begin{align*}
&\left| \sum_{s=1}^{k'} \sum_{t=1}^{k'-s+1} A_s B_t \right|_{ij}\\
& = \left| \sum_{s=2}^{k'+1} \sum_{i_2+j_2=s} A_{i_2} B_{j_2} \right|_{ij} \\
&\le \left( \sum_{s=2}^{k'+1} \sum_{i_2+j_2=s} |A_{i_2}| |B_{j_2}| \right)_{ij} \\
& \le \sum_{s=2}^{k'+1} \sum_{i_2+j_2=s} n 2^{-(i_2+j_2-2)\beta+2} g_i f_j \\
&= \sum_{s=2}^{k'+1} \sum_{i_2+j_2=s} n 2^{-(s-2)\beta+2} g_i f_j \\
&= \sum_{s=2}^{k'+1} (s-1) n 2^{-(s-2)\beta+2} g_i f_j.
\end{align*}
If $\beta \ge 3$, 
\[
\sum_{s=3}^{k'+1} (s-1) n 2^{-(s-2)\beta+2} g_i f_j \le  n 2^{2} g_i f_j
\]
Finally, for $\beta \ge 3$, we have
\begin{equation}
\left| \sum_{s=1}^{k'} \sum_{t=1}^{k'-s+1} A_s B_t \right|_{ij} \le 8n \cdot g_i f_j
\label{eq:bound}    
\end{equation}
From~\eqref{eq:multiple}, ~\eqref{eq:bound}, and ~\eqref{eq:no_error_condition}, if 
\begin{equation}
	2u \cdot \mathtt{ufp}(8n) \le 2^{-\beta (k'+1)+2}
	\label{eq:error-free-sum}
\end{equation}
is satisfied, then \eqref{eq:no_error_sum} holds.

% Let $s$ be the maximum value satisfying
% \[
% \lceil 14 + \log_2 n \rceil + \frac{1}{2}s(s-1) \le 53 -1.
% \]
Let $k'_{\max}$ be the maximum $k'$ satisfying \eqref{eq:error-free-sum}.
No rounding error occurs for $i=2,3,\dots,k'_{\max}$.
Therefore, we have
\begin{equation}
\begin{aligned}
&\left| \sum_{k=1}^k \sum_{j=1}^{k-i+1} A_i B_j - T_k \right|\\
&\le \left( \frac{1}{2} k (k+1) - \frac{1}{2} k'_{\max} (k'_{\max}+1) -1 \right) u |A| |B|.
\end{aligned}
\label{eq:Ootomo_sum}
\end{equation}
Summarizing, from~\eqref{eq:truncation} and~\eqref{eq:Ootomo_sum}, we have
\[
\begin{aligned}
&|AB - T_k| \\
&\le 4(k+1)n 2^{-\beta k} gf^T \\
&\quad + \left( \frac{1}{2} k (k+1) - \frac{1}{2} k'_{\max} (k'_{\max}+1) -1 \right) u |A| |B|.
\end{aligned}
\]
If \eqref{eq:well_conditioned} is satisfied, 
then we have
\[
\begin{aligned}
&|AB - T_k| \\
&\lesssim (k+1) 2^{- \beta k} |A||B| \\
&\quad + \left( \frac{1}{2} k (k+1) - \frac{1}{2} k'_{\max} (k'_{\max}+1) -1 \right) u |A| |B|.
\end{aligned}
\]

\subsection{Error bound for Algorithms~\ref{alg:OzakiSplit-ootomo} and \ref{alg:errfreesum}}

We focus on
\begin{equation}
\sum_{i_2+j_2=s} A_{i_2}'' B_{j_2}'' , \quad A_{i_2}'' \in \mathbb{I}_{\beta+1}^{m \times n}, \quad B_{j_2}'' \in \mathbb{I}_{\beta+1}^{n \times p}.
\label{eq:int8}
\end{equation}
We consider $r$ such that the following holds
\begin{equation}
\sum_{\substack{{i_2+j_2=s}\\ s \le \min(r,k+1)}} A_{i_2}'' B_{j_2}'' \in \mathbb{I}_{32}^{m \times p}.
\label{eq:nooverflow}   
\end{equation}
Let $E_{m,n}$ be the matrix $\mathbb{I}_2^{m \times n}$ whose elements are all ones.
From the definitions in~\eqref{eq:int8}, 
\[
|A_{i_2}''| \le (2^\beta-1) E_{m,n}, \quad |B_{i_2}''| \le (2^\beta-1)
E_{n,p},
\]
and we have
\begin{align*}
\left| \sum_{i_2+j_2=s} A_{i_2}'' B_{j_2}'' \right| 
& \le \sum_{i_2+j_2=s} | A_{i_2}''| |B_{j_2}''| \\
& \le (s-1)n (2^\beta-1)^2 E_{m,p}.
\end{align*}
Since the largest number in $\mathbb{I}_{32}$ is $2^{31}-1$, 
if $s \le r$ for $r := \max(1,2^{31 - 2\beta  - \lceil\log_2 n \rceil})$ in \eqref{eq:def_r},
we have
\begin{align*}
(s-1)n (2^\beta-1)^2
&\le (r - 1)n (2^\beta-1)^2\\
&\le (r-1)\cdot 2^{\lceil \log_2 n \rceil} \cdot 2^{2\beta}\\
&= r\cdot 2^{\lceil \log_2 n \rceil} \cdot 2^{2\beta} - 2^{\lceil \log_2 n \rceil} \cdot 2^{2\beta}\\
&\le 2^{31} - 1,
\end{align*}
which implies that \eqref{eq:nooverflow} holds.
Note that 
\[
(s-1)n (2^\beta-1)^2 < 2^{31} - 1
\]
for $\beta \ge 1$.
This is why constant $r$ was set as in \eqref{eq:def_r}.

If we define the number of the terms, $w$, for the accumulation as
\[
w:=\left\lceil \frac{k}{r} \right\rceil \left( k - \frac{r}{2} \left\lfloor \frac{k-1}{r} \right\rfloor \right), 
\]
we have
\[
|AB - T| \le 4(k+1)n 2^{-\beta k} gf^T + (w-1) u |A| |B|.
\]
If \eqref{eq:well_conditioned} is satisfied, 
then we have
\[
|AB - T| \lesssim (k+1) 2^{-\beta k} |A||B| + (w-1) u |A| |B|.
\]

\section{Conclusion}
\label{sec:Conclusion}
We proposed three implementation methods for accelerating the Ozaki scheme using the INT8 Tensor Core.
In the original implementation, \texttt{ozIMMU}, provided by Ootomo et al., the ratio of the accumulation in FP64 is not negligible, as it accounts for approximately 40--50 \% of the total computation time.
The proposed methods \texttt{ozIMMU\_EF} and \texttt{ozIMMU\_H} reduce the computation time ratio of the accumulation to approximately 10--20 \% and achieve a 1.2- to 1.6-fold speedup.
With future architectures expected to achieve high speed in lower-precision matrix multiplication, the computation time ratio of the accumulation in FP64 is expected to increase.
Thus, these proposed methods contribute to leveraging the performance of future architectures more effectively than \texttt{ozIMMU}.
The proposed methods \texttt{ozIMMU\_RN} and \texttt{ozIMMU\_H} offer alternative splitting methods using floating-point arithmetic in round-to-nearest-even mode and produce more accurate results than \texttt{ozIMMU} for the same number of slices.

\begin{acks}
We thank Dr. Hiroyuki Ootomo from NVIDIA for his helpful comments on the implementation of ozIMMU.
\end{acks}

\begin{dci}
The authors declare no competing interests.
\end{dci}

\begin{funding}
This study was partially supported by JSPS Grant-in-Aid for JSPS Fellows No. 22KJ2741, JSPS Grant-in-Aid for Research Activity Start-up No. 24K23874, and JSPS KAKENHI Grant No. 23K28100.
\end{funding}

\begin{sm}
 Not applicable.
\end{sm}

\bibliographystyle{SageH}
\bibliography{references.bib}

\begin{biogs}
Yuki Uchino is a postdoctoral researcher at RIKEN R-CCS. 
He received his Ph.D. in engineering from Shibaura Institute of Technology in 2024. 
His research interests include reliable computing, numerical linear algebra, and highly accurate algorithms. 
He won the student poster presentation award at the 38th JSST Annual International Conference on Simulation Technology (JSST 2019) and the student presentation awards at JSST 2021 and International Workshop on Reliable Computing and Computer-Assisted Proofs (ReCAP 2022).

Katsuhisa Ozaki is a full professor in the Department of Mathematical Sciences at Shibaura Institute of Technology. He received his Ph.D. in engineering from Waseda University in 2007. He was an Assistant Professor (2007-2008) and a Visiting Lecturer (2008-2009) at Waseda University. At Shibaura Institute of Technology, he has served as an Assistant Professor (2010-2013) and an Associate Professor (2013-2019), and has currently been a Professor since 2019. His research interests include reliable computing, particularly addressing rounding error problems in finite-precision arithmetic. He mainly focuses on numerical linear algebra and develops fast and accurate algorithms.

Toshiyuki Imamura is a team leader of Large-scale Parallel Numerical Computing Technology Team at RIKEN R-CCS, and is responsible for developing numerical libraries on Fugaku. He received his Diploma and Doctorate in Applied Systems and Sciences from Kyoto University in 1993 and 2000. He was a Researcher at CCSE, JAERI (1996-2003), a visiting scientist at HLRS (2002), and an associate professor at the University of Electro-Communications (2003-2012). His research interests include HPC, auto-tuning technology, and parallel eigenvalue computation. His research group won the HPL-MpX ranking (2020-2021) and was nominated as the Gordon Bell Prize finalist in SC05, SC06, and SC20.
\end{biogs}

\end{document}